\documentclass{article}
\usepackage{amsmath}
\usepackage{longtable}
\usepackage{graphicx}
\usepackage{amssymb}
\usepackage{amsmath}
\usepackage{graphicx}

\begin{document}

\begin{center}
\bigskip \textbf{MASS DECOMPOSITION\ OF SLACS LENS GALAXIES IN WEYL
CONFORMAL GRAVITY}

\bigskip

\bigskip

\bigskip

Alexander A. Potapov$^{1,a}$, Ramil N. Izmailov$^{2,b}$ and Kamal K. Nandi$^{1,2,3,c}$

$\bigskip $

$^{1}$Department of Physics \& Astronomy, Bashkir State University,
Sterlitamak Campus, Sterlitamak 453103, RB, Russia

$^{2}$Zel'dovich International Center for Astrophysics, M. Akmullah Bashkir
State Pedagogical University, Ufa 450000, RB, Russia

$^{3}$Department of Mathematics, University of North Bengal, Siliguri 734
013, India

\bigskip

$^{a}$E-mail: potapovaa2008@rambler.ru

$^{b}$E-mail: izmailov.ramil@gmail.com

$^{c}$E-mail: kamalnandi1952@yahoo.co.in

-------------------------------------------------------

\bigskip

\textbf{Abstract}
\end{center}

We study here, using the Mannheim-Kazanas solution of Weyl conformal theory,
the mass decomposition in the representative subsample of $57$ early-type
elliptical lens galaxies of the Sloan Lens Advanced Camera for Surveys
(SLACS) on board the Hubble Space Telescope (HST). We begin by showing that
the solution need not be an exclusive solution of conformal gravity but can
also be viewed as a solution of a class of $f(R)$ gravity theories coupled
to non-linear electrodynamics thereby rendering the ensuing results more
universal. Since lensing involves light bending, we shall first show that
the solution adds to Schwarzschild light bending caused by the luminous mass
($M_{\ast }$) a \textit{positive} contribution $+\gamma R$ contrary to the
previous results in the literature, thereby resolving a long standing
problem. The cause of the error is critically examined. Next, applying the
expressions for light bending together with an input equating Einstein and
Weyl angles, we develop a novel algorithm for separating the luminous
component from the total lens mass (luminous+dark) within the Einstein
radius. Our results indicate that the luminous mass estimates differ from
the observed total lens masses by a linear proportionality factor across the
subsample, which qualitatively agrees with the common conclusion from a
number of different simulations in the literature. In quantitative detail,
we observe that the ratios of luminous over total lens mass ($f^{\ast }$)
within the Einstein radius of individual galaxies take on values near unity,
many of which remarkably fall inside or just marginally outside the
specified error bars obtained from a simulation based on the Bruzual-Charlot
stellar population synthesis model together with the Salpeter Intitial Mass
Function (IMF) favored on the ground of metallicity 
[Grillo \textit{et al.}, Astron. Astrophys. \textbf{501}, 461(2009)].
We shall also calculate the average dark matter density 
$\left\langle \rho \right\rangle _{\text{av}}$ of individual galaxies within
their respective Einstein spheres. To our knowledge, the present approach,
being truly analytic, seems to be the first of its kind attempting to
provide a new decomposition scheme distinct from the simulational ones.


PACS no(s): 95.35.+d; 98.52.Eh; 98.62.Ck

\section{\textbf{Introduction}}

Unlike for spiral galaxies, which appear to be embedded in large dark matter
\textquotedblleft halos\textquotedblright , we cannot in general measure
rotation curves for elliptical galaxies, a vast majority of which act as
strong gravitational lenses. Only very rarely have strong lenses been
identified with spiral galaxies since the observed lens properties suggest
that they are produced not by the galactic disk but solely by the elliptical
galaxy bulge. Thus, using gravitational lensing, one can study the
mass-to-light ratio of elliptical galaxies, which shows that there is no
sign of large amount of dark matter surrounding these galaxies. If dark
matter is present in these galaxies, it has to be mixed in with the luminous
matter giving a total enclosed lens mass. We shall be concerned in this work
with the decomposition of this lens mass into dark and luminous parts within
the Einstein radius of elliptical galaxies.

A few lines about the dark matter hypothesis seems to be in order. Early
observations [1,2] on rotational data of spiral galaxies, now reconfirmed by
several observations extending well beyond the optical disc [3-14], indicate
that they do not conform to Newtonian gravity predictions. Doppler emissions
from stable circular orbits of neutral hydrogen clouds in the halo allow
measurement of tangential velocity $v_{\text{tg}}$ of the clouds treated as
probe particles. Contrary to Newton's laws, where $v_{\text{tg}}^{2}$ should
decay with radius $r$, observations indicate that it approximately levels
off with $r$ in the galactic halo region, which in turn calls for the
presence of additional non-luminous mass, the so called "dark matter".
Several well known theoretical models for dark matter exist in the
literature and it is impossible to list all of them here (only some are
mentioned in [15-39]). The most recent model, to our knowledge, seems to
be the so called Eddington-inspired Born-Infeld (EiBI) theory, succintly
called the "gravitational avatar of non-linear electrodynamics" [40],
developed by Ba\~{n}ados and Ferreira [41]. This new, and more general,
theory has led to many interesting observable predictions about dark matter
including the possibility of nonsingular cosmological models alternative to
inflation [42-49].

However, there is yet another variety of theories that do not at all require
dark matter for the interpretation of the observed rotation curves
associated with spirals. This class of theories include, e.g., Modified
Newtonian Dynamics (MOND) developed by Milgrom [50-52], Bekenstein and
Milgrom [53]. (The theory is very widely discussed, see e.g., [54,55]),
other modified gravities such as developed in [56], $f(R)$ gravity theories
[57,58], and Weyl conformal gravity\footnote{%
There has been much debate for and against the Weyl conformal gravity. For
instance, Flanagan [58] argues that if the source has associated with it a
macroscopic long range scalar field breaking conformal symmetry, the theory
does not reproduce attractive gravity in the solar system. However,
subsequently, Mannheim [60] has counter-argued that Schwarzschild tests of
solar gravity could still be recovered even in the presence of such
macroscopic fields.}[61].

We shall consider below what we call the Mannheim-Kazanas (MK) vacuum
(meaning matter-free) solution [61,62] of Weyl conformal gravity applying it
to elliptical lens galaxies.\footnote{%
{}The solution is called here the MK solution for easy reference. Note that
it is distinguished from Schwarzschild-de Sitter (SdS) solution by an extra 
\textit{linear} term $\gamma r$ contributed exclusively by conformal gravity
and reminding us of Mach's principle (see [63]). However, we shall soon show
that the MK solution can occur in other theories such as in $f(R)$ gravity
as well (see Sec.2C). The solution reduces to the familiar de Sitter form $%
B(r)=1-kr^{2}$ in the limit $M=0$, $\gamma =0$ and to the SdS form under $%
\gamma =0$.} The solution has three universal constant parameters, $\gamma
^{\ast }$, $\gamma _{0}$ and $k$, that are associated with potentials of
cosmic origin. This implies that the Weyl vacuum is not really empty but is
an arena for the energetic interplay of these potentials.\footnote{%
The approach involving potentials, though not necessarily of cosmic origin,
is not new. A potential determined by rotation curves, to our knowledge, was
first envisaged by Lake [23] that led to strong constraints on dark matter.
Conversely, a modified gravitational potential $\Phi $ fitting rotation
curves\ has been suggested by Capozziello \textit{et al} [64] in the context
of $f(R)$ gravity. The most commonly known potential is Newtonian, whose
strength $M_{\ast }$ dictates the deflection of light rays grazing the sun
in the environment of the Schwarzschild vacuum. Likewise, the Weyl vacuum
has two potentials $V_{\gamma _{0}}(r)$ and $V_{k}(r)$ making up the MK
metric and the associated potential strengths $\gamma _{0}$ and $k$ dictate
the light deflection there.} To be specific, the constant $\gamma _{0}$ is
associated with a universal linear potential term $V_{\gamma _{0}}(r)=\gamma
_{0}c^{2}r/2$ that is induced by the cosmic background and $k$ is associated
with a de Sitter-like potential term $V_{k}(r)=-kc^{2}r^{2}/2$ that is
induced by inhomogeneities in the cosmic background. The value $\gamma
^{\ast }$ is associated with the linear potential of the Sun and is so small
that there are no modifications to standard solar system phenomenology. The
constants were used to successfully fit the rotation data of individual
spiral galaxies including possible non-circular motions in the halo [65,66]
and also to determine their halo sizes from the condition of stability of
circular orbits [67]. However, we are not considering rotation curves in
this paper but nevertheless using those universal constants for the mass
decomposition in lens elliptical galaxies.

The purpose of the present paper is two-fold: First, we show that the light
bending in the MK solution enhances the Schwarzschild bending by an amount $%
+\gamma R$ contrary to the previous result of $-\gamma R$ in the literature.
We shall point out the causes for this discrepancy. The relative
contributions to bending from different terms will also be worked out.
Second, we shall find an application of this light bending. Using it in the
lens equation together with a certain logical input (explained below), we
shall investigate how far the MK solution can account for the mass
decomposition in the $57$ SLACS early-type elliptical lens galaxies [68],
which provide an unbiased subsample representative of the complete sample of
early-type galaxies in the Sloan Digital Sky Survey (SDSS) data base of over 
$10^{5}$ galaxies. This part of the task means that we shall be trying to
quantify the effect of the cosmology induced potentials $V_{\gamma _{0}}$
and $V_{k}$ on the galactic matter distribution within the Einstein radius
observed in lensing measurements.

To reach our goals, we shall calculate, following Ishak \textit{et al.}[69],
different contributions to light deflection in the MK solution using the
artefact of a vacuole that is assumed to enclose the galaxy (lens) at its
center. The predecessor of the vacuole method is the Rindler-Ishak method
[70], developed for asymptotically non-flat Schwarzschild-de Sitter (SdS)
metric of General Relativity (GR), that exposed the effect of cosmological
constant $\Lambda $ on light bending thereby debunking a prevailing belief
to the contrary [71]. Then we shall use a logical input, already employed in
the literature [48,49,72], viz., that the observed value of the Einstein
angle $\theta _{\text{E}}$ (caused by the luminous + dark matter) should be
equal to the Weyl angle $\theta _{\text{W}}$ (caused by the luminous matter
+ potentials), i.e., $\theta _{\text{Ein}}=\theta _{\text{Weyl}}$ for a
light ray having the \textit{same} impact parameter.\footnote{%
We wish to recall that Einstein and Weyl theories of gravity are both metric
theories but otherwise very different. Thus, by the equality $\theta _{\text{%
Ein}}=\theta _{\text{Weyl}}$, we are not saying that the two theories are
merging into one another as a whole but only saying that logically the
deflection must have a unique value for a unique impact parameter $b$, and
that all competing theories must predict the same numerical value. The true
justification for the adopted equality however has to come from other
observable \textit{predictions} that the equality would possibly lead to. In
fact, one such prediction is the galactic mass decomposition that can be
compared with those obtained by independent simulations available in the
literature [73].
\par
{}} Using the input, we shall develop a new algorithm that would lead to the
decomposition of the total lens mass $M_{\text{tot}}^{\text{lens}}$ within
the observed Einstein (or Weyl) radius into dark ($M_{\text{dm}}$) and
luminous ($M_{\ast }$) matter parts and compare the mass ratios with known
simulational predictions.

The paper is organized as follows: To get a reasonable view of what Weyl
conformal gravity looks like, we start in Sec.2 from the action discussing
certain pertinent issues related to conformal symmetry. We further show that
the MK metric can occur in the $f(R)$ gravity theory too thus making the
metric more universal than thought heretofore.\footnote{%
We thank an anonymous referee for suggesting this interesting possibility
and for advising some other major points relating to conformal symmetry to
be addressed in all the detail. Sec.2 is entirely devoted to these issues.}
In Sec.3, we provide critical reappraisals of some steps used previously in
the literature for the calculation of light deflection in in the MK metric.
Next, in Sec.4, we derive the light trajectory in the MK\ metric by
perturbatively solving the null geodesic equation and calculate in Sec.5 the
light deflection by using the vacuole method of Ishak \textit{et al.}[66].
In Sec.6, the algorithm for mass decomposition in the SLACS lens galaxies is
developed and applied to the considered subsample. The numerical results for
light bending and mass decomposition are shown in Tables I and II
respectively. Sec.7 concludes the paper. We shall choose units in which $G=1$%
, $\hslash =1$, the vacuum speed of light $c=1$ unless specifically
restored, and the signature chosen is ($+,-,-,-$).

\section{Weyl conformal gravity}

In recent times, it is increasingly realized that conformal gravity might
hold the key to resolving many outstanding problems of astrophysics. This
possibility has been advocated in a very recent article by 't Hooft [74] in
which he specifically refers to the works of MK including their 1989 paper
that brought to focus anew the conformal gravity.\footnote{%
A bit of curious history might be found in a recent article by Berezin 
\textit{et al} [63]: Conformal gravity was invented by Weyl in 1918 with a
motivation to combine the gravitational and electromagnetic fields into one
unified theory. However, the theory was rejected by Einstein and Weyl
because it was recognized that the conformal symmetry at most allows only
massless particles to exist. But this obstacle can be overcome today by
means of the Higgs mechanism for generating particle masses (for the latest
account, see [75]).} Further, it is well known that all the homogeneous and
isotropic space-times described by the Robertson-Walker line element have
zero Weyl tensor and are thus solutions to the vacuum conformal gravity.
Such a highly symmetric empty space-time is a good candidate for the
creation of the universe \textquotedblleft from nothing\textquotedblright\ $%
- $ a possibility first proposed by Vilenkin [76]. The idea that the initial
state of the universe should be conformally invariant was advocated also by
Penrose [77] and 't Hooft [74]. However, this empty universe might seem
unrealistic unless it could be filled with particles but how could this
happen? Spontaneous breaking of conformal symmetry in the early hot universe
giving rise to generation of particle masses is the answer (see the classic
works in [78-80]).

\textbf{A. The action and spontaneous breaking of conformal symmetry}

We start from the vacuum conformal gravity action 
\begin{equation}
I_{\text{W}}=-\alpha _{g}\int d^{4}x\sqrt{-g}C_{\alpha \beta \gamma \delta
}C^{\alpha \beta \gamma \delta },
\end{equation}%
where $\alpha _{g}$ is the coupling constant and $C^{\alpha \beta \gamma
\delta }$ is the Weyl tensor. This action is fully covariant with an
additional symmetry of conformal invariance under transformations of the
metric $g_{\mu \nu }(x)\rightarrow \Omega ^{2}(x)g_{\mu \nu }(x)$. Conformal
gravity thus possesses no fundamental scale (no intrinsic $G$ or fundamental 
$\Lambda $) at all, leading to an intrinsically scale-free cosmology at
sufficiently high enough temperatures [81]. Newton's constant $G_{N}$ might
be generated as a "macroscopic/low energy" limit (like the Fermi constant $%
G_{F}$ generated in the electroweak theory) to be measured by a Cavendish
experiment in a universe decoupled from the hot early stage.

We consider here conformal gravity exactly in this low energy limit, that
is, after spontaneous violation of symmetry has actually happened, particle
masses have been generated and galaxies formed as we see today. Conformal
gravity does not require elusive dark matter or dark energy for interpreting
(\textit{albeit} in a different way) the astrophysical observations,
provides singularity and ghost free solution to some of the known problems
plaguing standard cosmology\ including the cosmological constant [61,82] and
the age problem [83] (that is of course not to say that conformal gravity
has no problems of its own, though some seem to have been well answered, see
e.g., [60,84])

Variation of $g_{\mu \nu }$ in (1) leads to the field equations [61]%
\begin{equation}
4\alpha _{g}W_{\mu \nu }=T_{\mu \nu },
\end{equation}%
where%
\begin{equation}
W_{\mu \nu }\equiv W_{\mu \nu }^{(2)}-\frac{1}{3}W_{\mu \nu }^{(1)},
\end{equation}%
\begin{eqnarray}
W_{\mu \nu }^{(2)} &=&\frac{g_{\mu \nu }}{2}\nabla ^{\beta }\nabla _{\beta
}\left( R_{\alpha }^{\alpha }\right) +\nabla ^{\beta }\nabla _{\beta }R_{\mu
\nu }-\nabla _{\beta }\nabla _{\nu }R_{\mu }^{\beta }  \notag \\
&&-\nabla _{\beta }\nabla _{\mu }R_{\nu }^{\beta }-2R_{\mu \beta }R_{\nu
}^{\beta }+\frac{g_{\mu \nu }}{2}R_{\alpha \beta }R^{\alpha \beta }, \\
W_{\mu \nu }^{(1)} &=&2g_{\mu \nu }\nabla ^{\beta }\nabla _{\beta }\left(
R_{\alpha }^{\alpha }\right) -2\nabla _{\mu }\nabla _{\nu }R_{\alpha
}^{\alpha }-2R_{\alpha }^{\alpha }R_{\mu \nu }+\frac{g_{\mu \nu }}{2}R^{2}.
\end{eqnarray}%
($\nabla _{\beta }$ is the covariant derivative operator) and $T_{\mu \nu }$
is the conformally invariant energy momentum tensor to be supplied. Solving
these intimidating system of equations, MK [62] computed the solution
exterior ($T_{\mu \nu }=0$) to a static, spherically symmetric gravitating
source, which is 
\begin{eqnarray}
d\tau ^{2} &=&B(r)dt^{2}-\frac{1}{B(r)}dr^{2}-r^{2}(d\theta ^{2}+\sin
^{2}\theta d\varphi ^{2}),\text{ \ } \\
B(r) &=&1-\frac{\beta (2-3\beta \gamma )}{r}-3\beta \gamma +\gamma r-kr^{2},
\end{eqnarray}%
where $\beta ,\gamma ,k$ are dimensionful integration constants. Defining
the Schwarzschild mass $M=\frac{\beta (2-3\beta \gamma )}{2}$, we can
rewrite the exact metric as 
\begin{equation}
B(r)^{\text{exact}}=\alpha -\frac{2M}{r}+\gamma r-kr^{2},
\end{equation}%
where $\alpha \equiv (1-6M\gamma )^{1/2}$. From the observed galaxy (treated
as lens) masses $M$ and the rotation curve fitted universal value of $\gamma 
$, the constant factor $6M\gamma \approx 10^{-15}$, and can be easily
neglected for simplicity. Thus, for the galactic mass decomposition, we
shall take the MK metric relevant on the scales of Einstein radius of lens
mass $M$ and beyond to be\footnote{%

The reason why the $B(r)$ are separately designated as exact and galactic is
that the metric (9) does not follow from (8) at the exact value $\alpha =1$
as it would then require either $M$ or $\gamma $ or both to be exactly zero.
However, one could technically say that (9) follows from (8) only in the 
\textit{limit} $\alpha \rightarrow 1$. Physically, it implies that the
metric (9) is can apply to a distance scale, where effects of both $M$ and $%
\gamma $ are non-zero, with the combined effect $M\gamma $ to be negligibly
small. Such a scale is naturally provided by the galactic halo radius [67],
which is intermediate between the Schwarzschild radius of $M$ and the
cosmological de Sitter radius.}%
\begin{equation}
B(r)^{\text{galactic}}=1-\frac{2M}{r}+\gamma r-kr^{2},
\end{equation}%
that has also been used for predicting flat rotation curves in the galactic
halo with universal values of $\gamma $ and $k$ in the galactic halo region
[65]. The emergence of Schwarzschild mass together with other dimensionful
constants in the metric (8) already indicates the first instance of local
symmetry violation in the vacuum Weyl gravity at the solution level. As a
second instance of symmetry violation, we consider conformal cosmology.

\textbf{B. Conformal cosmology}

Consider the action of the conformally coupled matter [81] 
\begin{eqnarray}
I_{\text{M}} &=&-\hslash \int d^{4}x\sqrt{-g}\left[ \underbrace{\frac{1}{2}%
\nabla ^{\mu }\phi \nabla _{\mu }\phi -\frac{1}{12}\phi ^{2}R+\lambda \phi
^{4}}+\underbrace{i\overline{\psi }\gamma ^{\mu }(x)[\partial _{\mu }+\Gamma
_{\mu }(x)]\psi }\right.  \notag \\
&&\text{ \ \ \ \ \ \ \ \ \ \ \ \ \ \ \ \ \ \ \ \ \ \ \ \ \ \ \ \ \ \ \ \ }%
scalar\text{\ \ \ \ \ \ \ \ \ \ \ \ \ \ \ \ \ \ \ \ \ \ \ }\ \ \ \ \
fermion\ \text{\ \ \ \ \ }  \notag \\
&&\left. -\underbrace{g\phi \overline{\psi }\psi }\right] \\
&&interaction  \notag
\end{eqnarray}%
where $\Gamma _{\mu }(x)$ is the fermion spin connection, $\lambda $ and $g$
are the dimensionless coupling constants, $\phi (x)$ is the symmetry
breaking scalar field and $\psi $ is a fermionic field. The matter energy
momentum tensor following from the action (10) is%
\begin{eqnarray}
T^{\mu \nu } &=&\hslash \lbrack i\overline{\psi }\gamma ^{\mu }(\partial
_{\mu }+\Gamma _{\mu })\psi +\frac{2}{3}\nabla ^{\mu }\phi \nabla ^{\nu
}\phi -\frac{g^{\mu \nu }}{6}\nabla ^{\alpha }\phi \nabla _{\alpha }\phi -%
\frac{\phi }{3}\nabla ^{\mu }\nabla ^{\nu }\phi  \notag \\
&&+\frac{g^{\mu \nu }\phi \nabla ^{\alpha }\nabla _{\alpha }\phi }{3}-\frac{%
\phi ^{2}}{6}(R^{\mu \nu }-\frac{1}{2}g^{\mu \nu }R)-g^{\mu \nu }\lambda
\phi ^{4}].
\end{eqnarray}%

Defining the "density" $\rho $ of a perfect fluid (with $u^{\mu }$ being the
time-like four-velocity satisfying $u^{\mu }u_{\mu }=1$) 
\begin{equation}
\rho u^{\mu }u^{\nu }=i\hslash \overline{\psi }\gamma ^{\mu }(\partial ^{\nu
}+\Gamma ^{\nu })\psi +\frac{\hslash }{2}\nabla ^{\mu }\phi \nabla ^{\nu
}\phi ,
\end{equation}%
and isotropic "pressure" $p$ 
\begin{equation}
pu^{\mu }u^{\nu }=-\frac{\hslash }{3}\phi \nabla ^{\mu }\nabla ^{\nu }\phi +%
\frac{\hslash }{6}\nabla ^{\mu }\phi \nabla ^{\nu }\phi ,
\end{equation}%
the energy momentum tensor may be rewritten in a more elegant form as 
\begin{equation}
T_{\mu \nu }=(p+\rho )u_{\mu }u_{\nu }+pg_{\mu \nu }-\frac{\phi ^{2}}{6}%
(R_{\mu \nu }-\frac{1}{2}g_{\mu \nu }R)-g_{\mu \nu }\lambda \phi ^{4}.
\end{equation}%

This provides for the right hand side of Eq.(2). In an isotropic and
homogeneous universe, the left hand side of Eq.(2) is identically zero
(empty universe), thus leading to $T_{\mu \nu }=0$. When the scalar field $%
\phi (x)$ in $I_{\text{M}}$ obtains a non-zero mass (which we are free to
rotate to some "spacetime constant" $\phi _{0}$ due to conformal freedom),
we get%
\begin{equation}
\frac{\phi _{0}^{2}}{6}(R_{\mu \nu }-\frac{1}{2}g_{\mu \nu }R)=(p+\rho
)u_{\mu }u_{\nu }+pg_{\mu \nu }-g_{\mu \nu }\lambda \phi _{0}^{4}.
\end{equation}%

Thus conformal cosmology looks like the standard cosmology with a "perfect
fluid" source and a non-zero cosmological constant $\Lambda =\lambda \phi
_{0}^{4}$ with the important exception that Newton's constant $G_{N}$ has
been replaced by an "effective" constant of the form 
\begin{equation}
G_{\text{eff}}=-\frac{3}{4\pi \phi _{0}^{2}},
\end{equation}%
just as has been advocated by 't Hooft [74]. This is not the low energy
Newton's constant $G_{N}$ that Cavendish measured, but instead a term which
we identify to be the negative gravitational constant $G_{\text{eff}}$
providing \textit{repulsion} at cosmological distances. Thus, in the
isotropic and homogeneous case, we end up breaking conformal symmetry again
recovering standard cosmology with a non-zero cosmological constant $\Lambda 
$. Local gravity is fixed by small, local variations in the background
scalar field $\phi (x)$, with such variations being completely decoupled
from the homogeneous, constant, cosmological background field $\phi _{0}$
itself. It is the distinction between inhomogeneity on the local scale and
homogeneity on the global scale that provides the demarcation between local
and global gravity respectively. Hence conformal gravity is attractive on
local galactic scales, while it is repulsive on cosmological scales, a fact
that has been explicitly demonstrated very recently by Phillips [85].

Finally, we wish to point out that the MK metric (8) or (9) differ from the
SdS black hole by an important linear term $\gamma r$, which is a specific
contribution from the fourth order vacuum conformal gravity. Nonetheless,
the metric need not correspond exclusively to vacuum conformal gravity. An
alternative theory of which the MK metric may again be a solution could be a
similar fourth order theory. The best candidates are the $f(R)$ gravity
theories that have been very widely discusssed in the literature (see the
review [86], [87-94]), notably in connection with modelling dark matter and
dark energy as curvature effects [96], instability and anti-evaporation of
black holes [95,96]. For a full account of these and other effects, we refer
the reader to the excellent treatise [97]. We shall now show that the MK
metrics (8,9) could indeed be viewed as a solution of $f(R)$ gravity coupled
to non-linear electrodynamics (NED), so that the results of the present
paper are actually more universal than thought heretofore, these now being
valid in a wider class of $f(R)$ theories as well.

\textbf{C. MK metric as }$f(R)$\textbf{\ gravity solution}

The $f(R)$ gravity is more general than GR, the latter recovered only when $%
f(R)=R$. To reach our goal stated above, we shall follow an algorithm
developed by Rodrigues \textit{et al }[98]. A similar algorithm was
developed much earlier by Capozziello \textit{et al} [99]. They have
recently obtained solutions of $f(R)$ gravity coupled to NED that may be
viewed as generalizations of known solutions within the GR theory. Here we
are considering a known solution of conformal gravity instead of GR. The
action for $f(R)$ gravity with sources is given by\ [99] 
\begin{equation}
I_{f(R)}=\int d^{4}x\sqrt{-g}\left[ f(R)+2\kappa ^{2}\mathcal{L}_{m}\right] ,
\end{equation}%
where $\kappa ^{2}=8\pi G/c^{4}$ and $\mathcal{L}_{m}$ is the matter
Lagrangian. Varying the action with respect to the metric yields the field
equations%
\begin{equation}
f_{R}R_{\nu }^{\mu }-\frac{1}{2}\delta _{\nu }^{\mu }f+\left( \delta _{\nu
}^{\mu }\square -g^{\mu \beta }\nabla _{\beta }\nabla _{\nu }\right)
f_{R}=\kappa ^{2}\Theta _{\nu }^{\mu },
\end{equation}%
where $f_{R}=df(R)/dR,\square \equiv g^{\alpha \beta }\nabla _{\alpha
}\nabla _{\beta }$ is the D'Alembertian and $\Theta _{\nu }^{\mu }$ is the
source stress tensor. With $\mathcal{L}_{m}=\mathcal{L}_{\text{NED}}(F)$,
where $F=\frac{1}{4}F^{\mu \nu }F_{\mu \nu }$, $F_{\mu \nu }=\partial _{\mu
}A_{\nu }-\partial _{\nu }A_{\mu }$ is the Faraday tensor [100], it follows
that $\Theta _{\nu }^{\mu }=\delta _{\nu }^{\mu }\mathcal{L}_{\text{NED}}-%
\frac{\partial \mathcal{L}_{\text{NED}}(F)}{\partial F}F^{\mu \alpha }F_{\nu
\alpha }.$ Varying $A_{\mu }$ in (17), one has $\nabla _{\mu }\left[ F^{\mu
\nu }\mathcal{L}_{\text{F}}\right] =0$, where the two Lagrangian densities
are related by $\mathcal{L}_{\text{F}}=\frac{\partial \mathcal{L}_{\text{NED}%
}(F)}{\partial F}=\frac{\partial \mathcal{L}_{\text{NED}}(F)}{\partial r}%
\left( \frac{\partial F}{\partial r}\right) ^{-1}$. Due to radial symmetry
of the metrics, the only non-zero component of the Faraday tensor is $%
F^{10}(r)=\frac{q}{r^{2}}\mathcal{L}_{\text{F}}^{-1}$ and $F=-\frac{1}{2}%
\left( F^{10}\right) ^{2}$. The field equations (18) can be rewritten in
terms of effective stress tensor of GR ("curvature fluid" [64]):%
\begin{equation}
R_{\mu \nu }-\frac{1}{2}g_{\mu \nu }R=f_{R}^{-1}\left[ \kappa ^{2}\Theta
_{\mu \nu }+\frac{1}{2}g_{\mu \nu }(R-Rf_{R})-(g_{\mu \nu }\square -\nabla
_{\mu }\nabla _{\nu })f_{R}\right] \equiv \kappa ^{2}T_{\mu \nu }^{\text{eff}%
}\text{.}
\end{equation}

Rodrigues \textit{et al }[98] start from a radial mass function $M(r)$
which, for the metric (9), is obtained by rewriting $B(r)=1-\frac{2M(r)}{r}$:%
\begin{equation}
M(r)=M\left( 1-\frac{\gamma r^{2}}{2M}+\frac{kr^{3}}{2M}\right) ,
\end{equation}%
where $M$ is the constant Schwarzschild mass from the metric (9). The Ricci
scalar $R(r)$ is independent of $M$:%
\begin{equation}
R=\frac{6}{r}\left( \gamma -2kr\right) \Rightarrow r(R)=\frac{6\gamma }{R+12k}.
\end{equation}%

We give below only the final results for the galactic scale metric (9).
Following the algorithm in [98], it follows that the class of $f(R)$
gravity theories 
\begin{equation}
f(R)^{\text{galactic}}=c_{0}R+c_{1}\int r(R)dR=c_{0}R+6c_{1}\gamma \ln
\left( R+12k\right) ,
\end{equation}%
where $c_{0}$ and $c_{1}$ are arbitrary constants, would also produce the
MK\ solution (9). The NED sector gives%
\begin{equation}
F^{10}=-\frac{2c_{0}\gamma r+c_{1}\left\{ r\left( 2+3\gamma r\right)
-6M\right\} }{2\kappa ^{2}q},
\end{equation}%
\begin{equation}
\mathcal{L}_{\text{F}}=-\frac{2\kappa ^{2}q^{2}}{r^{2}\left[ 2c_{0}\gamma
r+c_{1}\left\{ r\left( 2+3\gamma r\right) -6M\right\} \right] },\mathcal{L}_{%
\text{NED}}=-\frac{c_{0}\gamma +c_{1}\left( 1-3\gamma r\ln r\right) }{\kappa
^{2}r}.
\end{equation}

We see that $f(R)^{\text{galactic}}$ is a nice logarithmic function that can
be analyzed by choosing values of $c_{0}$ and $c_{1}$ and/or solution
constants $\gamma $ and $k$. For instance, $\gamma =0$ gives the SdS case
with $f(R)=R=-12k$. For $k=0$, one obtains $f(R)=c_{0}R+6c_{1}\gamma \ln R$,
which vividly shows the role of $f(R)$ beyond GR leading to the emergence of
an\ extra linear term $\gamma r$ in the metric (9), that have both Cauchy
and event horizons (see Sec.3), as a deviation from the SdS black hole. The
influence of $\gamma r$ on the anti-evaporation phenomenon will be a
challenging and interesting future task. Exactly the same algorithm applies
also to the exact form (8) yielding%
\begin{eqnarray}
f(R)^{\text{exact}} &=&c_{0}R+2c_{1}\left[ \sqrt{9\gamma
^{2}+2(R+12k)(\alpha -1)}\right.   \notag \\
&&\left. +3\gamma \ln \left( \sqrt{9\gamma ^{2}+2(R+12k)(\alpha -1)}-3\gamma
\right) \right] .
\end{eqnarray}%

The energy conditions, except the strong energy condition, are satisfied by
the $T_{\mu \nu }^{\text{eff}}$ for both the metrics (8) and (9), now posed
as solutions of $f(R)$ gravity. The Dolgov-Kawasaki [100] stability
condition $\frac{d^{2}f(R)}{dR^{2}}\geq 0$ is satisfied only when $c_{1}\leq
0$ in both (22) and (25). It is interesting to note that logarithmic $%
f(R)=\ln (\lambda R)$, investigated in [95], follow as special cases Eq.(22)
under $c_{0}=0$, $k=0$. A fuller analysis for both the $f(R)$ gravities
equivalent to conformal solutions will be given elsewhere as they will take
us outside the scope of the present paper.

\section{Critical reappraisals}

We shall here critically re-examine three issues that relate to light
deflection in the MK metric of Weyl gravity. Some years ago, Edery and
Paranjape [72] calculated the light deflection in the asymptotically
non-flat MK metric (9) to be $\Delta \phi =\frac{4M}{R}-\gamma R$, where $M$
is the luminous mass of a galaxy, $\gamma $ is a constant parameter and $R$
is identified with the closest approach distance $r_{0}$. However, Edery and
Paranjape already recognized that the negative sign in $-\gamma R$ was
discrepant requiring further investigation because, only for $\gamma <0$,
the contribution becomes positive imitating the effect of attractive dark
matter but then the problem is that rotation curve fit requires $\gamma >0$,
an exactly opposite sign.\ 

We argue here that the discrepant sign is a result of an illegitimate range
of integration. To derive the above deflection, Edery \& Paranjape
considered integration over the radial coordinate $r$ from $R$ to $\infty $
arguing that the incoming light followed a "straight line" path at infinity
associated with the metric function $B_{\infty }(r)=1+\gamma r-kr^{2}>0$, $%
k>0$ [neglecting $M/r$ in metric (9)]. It is this reduced metric that, under
transformation $r\rightarrow \rho $ [see Eq.(28) below], is conformal to
cosmological Robertson-Walker metric with negative space curvature ($K=-k-%
\frac{\gamma ^{2}}{4}$) [62]. They further argued that the "straight line"
path was justified by the limit $\frac{d\varphi }{dr}\rightarrow 0$ as $%
r\rightarrow \infty $. We wish to point out that the limit $r\rightarrow
\infty $ does not make sense because there is a finite horizon radius in the
metric, which limits the motion of light \textit{inside} this radius.
Outside the horizon, at $r\rightarrow \infty $, where the light ray has been
assumed to pass, the metric function $B_{\infty }(r)$ changes sign leading
to violation of the metric signature, which forbids a meaningful integration
from $R$ to $\infty $.

To be more specific, the metric function $B_{\infty }(r)=0$ gives the
horizon radius 
\begin{equation}
r=r_{\text{hor}}=\frac{\gamma +\sqrt{4k+\gamma ^{2}}}{2k}.
\end{equation}%

For the special case $\gamma =0$ and $k=\frac{\Lambda }{3}>0$, one simply
retrieves the de Sitter horizon radius $r_{\text{hor}}^{\text{dS}}=\sqrt{%
\frac{3}{\Lambda }}$. Outside the horizon $r>$ $r_{\text{hor}}$, say at $%
r=2r_{\text{hor}}$, $B_{\infty }(2r_{\text{hor}})$ becomes generically
negative no matter what the sign of $\gamma $ is.\footnote{%
We understand that "straight line" on a curved geometry is distinguished
here from the Euclidean straight line on flat geometry. However, the issue
is not about this distinction but about the validity of the limit $%
r\rightarrow \infty $. The situation here is exactly opposite to what one
finds, e.g., in the Schwarzschild geometry, where $B(r)=1-2M/r>0$ for $r>r_{%
\text{hor}}=2M$.} This can be seen from 
\begin{equation}
B_{\infty }(2r_{\text{hor}})=-\frac{3k+\gamma \left( \gamma +\sqrt{4k+\gamma
^{2}}\right) }{k}<0.
\end{equation}%

On the other hand, inside the horizon $r<$ $r_{\text{hor}}$, say at $r=r_{%
\text{hor}}/2$, $B_{\infty }(r_{\text{hor}}/2)>0$, which is consistent with
the required signature protection, thereby limiting the light motion to only
within $r<$ $r_{\text{hor}}$. If one changes from static $r$ to comoving
radial coordinate $\rho $ by [62] 
\begin{equation}
\rho (r)=\frac{4r}{2\left( 1+\gamma r-kr^{2}\right) ^{1/2}+2+\gamma r},
\end{equation}%
one obtains 
\begin{equation}
\rho _{\text{hor}}(r_{\text{hor}})=\frac{4\left( \gamma +\sqrt{4k+\gamma ^{2}%
}\right) }{4k+\gamma \left( \gamma +\sqrt{4k+\gamma ^{2}}\right) },
\end{equation}%
which is just the redefined horizon radius. However, it can be easily
verified that $\rho (r=2r_{\text{hor}})$ is imaginary! Also, as $%
r\rightarrow \infty $, $\rho \rightarrow \frac{4}{2\sqrt{-k}+\gamma }$,
which too is imaginary for $k>0$ and so is physically meaningless.

Exactly the same arguments hold for the full MK solution $B(r)=1-\frac{2M}{r}%
+\gamma r-kr^{2}$ as well because there now appears two horizons for $k>0$
at radii given by (for details, see Ref.[101]): 
\begin{eqnarray}
r_{\text{hor}}^{1} &=&2\sqrt{\frac{3k+\gamma ^{2}}{9k^{2}}}\cos \left( \frac{%
\theta +4\pi }{3}\right) +\frac{\gamma }{3k}\text{, } \\
r_{\text{hor}}^{2} &=&2\sqrt{\frac{3k+\gamma ^{2}}{9k^{2}}}\cos \left( \frac{%
\theta }{3}\right) +\frac{\gamma }{3k}\text{, } \\
r_{\text{hor}}^{2} &>&r_{\text{hor}}^{1}\text{, }\theta =\theta (M,\gamma
,k),
\end{eqnarray}%
where $r_{\text{hor}}^{1}$ and $r_{\text{hor}}^{2}$ are the radii of the
inner Cauchy horizon and the outer event horizon respectively and away
outward from the event horizon, the metric function $B(r)$ becomes negative.
Thus, due to the presence of horizons, the usual integration from $R$ to $%
\infty $ does not make sense either in the mutilated version ($M=0$) or in
the full MK solution ($M\neq 0$). Hence, the resultant deflection $\Delta
\phi =\frac{4M}{R}-\gamma R$ cannot be accepted as valid. To find the valid
light deflection, it is necessary to find a new method that does not rely on
such integration. That very new method has been developed by Rindler and
Ishak [70], which is based on the geometric invariant angle and is most
suited to the asymptotically non-flat spacetimes. We shall soon see that the
bending expression, apart from the term $+\gamma R$, yields also other terms
in which $\gamma $ couples with $M$, $k$ and $R$ in different combinations.

The second issue relates to two previous papers [102,103] that unfortunately
overlooked a practical condition, viz., that the closest approach distance $%
R $ must be far greater than the Schwarzschild radius of the galaxy, i.e., $%
R\gg 2M$. This condition technically induces a certain function $\left\vert
A\right\vert $ to assume a \textit{positive} value, which is crucial for
obtaining the known expression for bending. For instance, with $r=1/u$,
Rindler and Ishak [70] defined the function $A(r,\varphi )\equiv \frac{dr}{%
d\varphi }=\left( -r^{2}\right) \frac{du}{d\varphi }$ but used only the
positive numerical value $\left\vert A\right\vert =r^{2}\frac{du}{d\varphi }=%
\frac{R^{3}}{4M^{2}}$ in their prescription for bending that led to the
correct Schwarzschild deflection. The importance of their positivity
prescription is that, without it even the known Schwarzschild deflection
would not follow. Since this positivity was not accommodated in [102,103],
an erroneous two-way negative contribution $-\gamma R$ appeared there,
supporting the existing result of Edery and Paranjape [72]. We shall respect
this prescription in the present paper, which will show that the two-way
contribution in Weyl gravity actually is $+\gamma R$ thus enhancing the
Schwarzschild bending and imitating the effects of attractive dark matter,
as should be the case.

The third and final issue concerns the appropriate set up needed for
calculating the light deflection. Note that the Rindler-Ishak method
originally proposed in [70] was based on a source and an observer located in
a static SdS background. On the other hand, given the environment of
Mannheim-O'Brien [65] cosmological potentials, it should be more appropriate
to derive the corresponding light bending equation in a cosmological set up,
that is, in the Friedmann-Lema\^{\i}tre-Robertson-Walker (FLRW) background.
To do that, Ishak \textit{et al.} [69] extended the original method in which
the galaxy (lens) is now placed at the center of a SdS vacuole exactly
embedded into the FLRW spacetime using the Einstein-Strauss [104]
prescription and appropriate junction conditions. We shall follow this
extended method in this paper.

\section{Light trajectory in the MK spacetime}

We restate the asymptotically non-flat spherically symmetric galactic MK
solution as used in [65,72]:%
\begin{equation}
d\tau ^{2}=B(r)dt^{2}-\frac{1}{B(r)}dr^{2}-r^{2}(d\theta ^{2}+\sin
^{2}\theta d\varphi ^{2}),\text{ \ }B(r)=1-\frac{2M}{r}+\gamma r-kr^{2},%
\text{\ }
\end{equation}%
where $M$ is the luminous central mass, $k$ and $\gamma $ are constants.
Denoting $u=1/r$, we derive the path equation for a test particle of mass $%
m_{0}$ on the equatorial plane $\theta =\pi /2$ as follows:%
\begin{equation}
\frac{d^{2}u}{d\varphi ^{2}}+u=3Mu^{2}-\frac{\gamma }{2}+\frac{M}{h^{2}}+%
\frac{1}{2h^{2}u^{2}}\left( \gamma -\frac{2k}{u}\right) ,
\end{equation}%
where $h=\frac{J}{m_{0}}$, the conserved angular momentum per unit test
mass. For photon, $m_{0}=0\Rightarrow h\rightarrow \infty $ and one ends up
with the null geodesic equation:%
\begin{equation}
\frac{d^{2}u}{d\varphi ^{2}}+u=3Mu^{2}-\frac{\gamma }{2}.
\end{equation}%

We shall perturbatively solve this equation. To zeroth order, Eq.(35) gives%
\begin{equation}
\frac{d^{2}u_{0}}{d\varphi ^{2}}+u_{0}=-\frac{\gamma }{2}
\end{equation}%
and its exact solution is%
\begin{equation}
u_{0}=\frac{\cos \varphi }{R}-\frac{\gamma }{2}
\end{equation}%
where $R$ is a constant related to the distance of closest approach $r_{0}$
to the origin. For transparency, we shall consider only first order
perturbation in $M$. Thus, following the usual method of small
perturbations, we want to derive the solution as%
\begin{equation}
u=u_{0}+u_{1}
\end{equation}%
where $u_{1}$ satisfies%
\begin{equation}
\frac{d^{2}u_{1}}{d\varphi ^{2}}+u_{1}=3Mu_{0}^{2}.
\end{equation}

The perturbative expansion holds only for small $u$ or large $r$. Thus we
are considering galactic parameters $M$, $R$ and solution parameter $\gamma $
such that the non-dimensional quantities $\frac{2M}{R}\ll 1$ and $\gamma
R\ll 1$. The exact solution of Eq.(39) is%
\begin{equation}
u_{1}=\frac{M}{4R^{2}}\left[ 6+3R^{2}\gamma ^{2}-6R\gamma \cos \varphi
-2\cos 2\varphi -6R\gamma \varphi \sin \varphi \right] .
\end{equation}%

When $\gamma =0$, it may be verified that one recovers the equation for
light trajectory in first order in the Schwarzschild metric. Formally
changing $\varphi \rightarrow \frac{\pi }{2}-\varphi $, the final solution
for light trajectory up to first order in $M$ can be written as%
\begin{align}
u& \equiv \frac{1}{r}=\frac{\sin \varphi }{R}-\frac{\gamma }{2}  \notag \\
& +\frac{M}{4R^{2}}\left[ 6+3R^{2}\gamma ^{2}-3R\gamma (\pi -2\varphi )\cos
\varphi +2\cos 2\varphi -6R\gamma \sin \varphi \right] .
\end{align}%

The closest approach distance $r_{0}$ is obtained from Eq.(41) by putting $%
\varphi =\pi /2$ and is given by%
\begin{equation}
\frac{1}{r_{0}}=\frac{1}{R}+\left[ \frac{M(4-6R\gamma +3R^{2}\gamma ^{2})}{%
4R^{2}}-\frac{\gamma }{2}\right] \simeq \frac{1}{R},
\end{equation}%
because for typical observed galactic values of $M,R$ and $\gamma $, that we
shall soon see, it follows that $\frac{1}{R}\approx 10^{-22}$ cm$^{-1}$,
while the piece in the square bracket $\approx 10^{-30}$ cm$^{-1}$, hence it
can be ignored\footnote{%
The symbol "$\approx $" \ means "of the order of".}. Thus, $R$ \ can be
identified with the closest approach distance: 
\begin{equation}
R\simeq r_{0}.
\end{equation}%

Eq.(41) is the desired equation to be used in the sequel.

\section{Light deflection:\ Ishak \textit{et al.} vacuole method}

The original method proposed by Rindler \& Ishak [70] did not require the
concept of a vacuole because the source and observer were assumed to be
located in a static SdS background. The work nonetheless contained the
essential ingredient for its vacuole extension: It combined the standard
perturbative solution with an invariant geometric definition of the bending
angle that took into account the explicit effect of the metric $B(r)$ that
contains $k$, which is formally similar to but not numerically exactly the
same as $\Lambda $.

The invariant\ geometric formula for the cosine of the angle $\psi $ between
two coordinate directions $d$ and $\delta $ is given by%
\begin{equation}
\cos \psi =\frac{g_{ij}d^{i}\delta ^{j}}{(g_{ij}d^{i}d^{j})^{1/2}(g_{ij}%
\delta ^{i}\delta ^{j})^{1/2}}.
\end{equation}%

Differentiating $u$ with respect to $\varphi $, denoting $\frac{dr}{d\varphi 
}=\left( -r^{2}\right) \frac{du}{d\varphi }\equiv A(r,\varphi )$, and
imposing the Rindler-Ishak positivity condition on $A(r,\varphi )$ [67], we
get%
\begin{equation}
\left\vert A(r,\varphi )\right\vert =r^{2}\frac{du}{d\varphi }.
\end{equation}%

For the light path Eq.(41), the function $\left\vert A\right\vert $ reads: 
\begin{equation}
\left\vert A(r,\varphi )\right\vert =\left( r^{2}\right) \times \left[ \frac{%
3MR\gamma (\varphi -2\pi )\sin \varphi +4(R-2M\sin \varphi )\cos \varphi }{%
4R^{2}}\right] >0.
\end{equation}

Ishak \textit{et al.} [69] method treats the light bending in a cosmological
setting that requires the concept of a vacuole, which is not present in
nature by itself but devised here only as an artefact of the investigative
procedure. The vacuole is assumed to be a large hypothetical sphere that
houses the lens galaxy at its center and is exactly embedded into the FLRW
spacetime. It is further assumed that all the light-bending occurs inside
the vacuole and that once the light transitions out of the vacuole and into
FLRW spacetime, all bending stops. The cut-off point, where the transition
occurs, is tailored to each individual lens, namely the edge of the vacuole
defined by its radius $r_{b}$. The vacuole concept as such is not used
except for this one purpose, viz., to obtain a realistic order-of-magnitude
estimate of the range of influence of the lens. Once we are able to obtain
the deflection angle, we can dismiss the vacuole as redundant.

Assuming a small light entry angle $\varphi _{b}$ at the vacuole boundary
radius $r=r_{b}$ such that $\sin \varphi _{b}\simeq \varphi _{b}$, and $\cos
\varphi _{b}\simeq 1$, Eq.(41) gives 
\begin{equation}
\frac{1}{r_{b}}=\frac{\varphi _{b}}{R}+\frac{2M}{R^{2}}+\left[ M\left( \frac{%
3\gamma ^{2}}{4}-\frac{3\pi \gamma }{4R}\right) -\frac{\gamma }{2}\right] ,
\end{equation}%
or, equivalently%
\begin{equation}
\varphi _{b}=\left( \frac{R}{r_{b}}-\frac{2M}{R}\right) +\left[ \frac{3\pi
M\gamma }{4}+\frac{\gamma R}{2}-\frac{3MR\gamma ^{2}}{4}\right] .
\end{equation}

Note that we have only one equation (47) connecting two unknowns $\varphi
_{b}$ and $r_{b}$. Hence we need to specify any one of them from independent
considerations. Ishak \textit{et al.}[69] employed the Einstein-Strauss
prescription [104,105] to determine the boundary $r_{b}$ assuming that the
SdS vacuole has been matched to an expanding FLRW universe via the
Sen-Lanczos-Darmois-Israel junction conditions [106-109]. In general, the
vacuole radius $r_{b}$ would also change due to cosmic expansion, but Ishak 
\textit{et al.}[69] considered $r_{b}$ at that particular instant $t_{0}$ of
cosmic epoch, when the light ray just happens to pass the point of closest
approach to the lens. The Einstein-Strauss prescription adapted to the MK
solution reads: 
\begin{equation}
r_{b\text{ in MK}}=a(t)r_{b\text{ in FLRW}}\text{, }M_{\text{MK}}=\frac{4\pi 
}{3}r_{b\text{ in MK }}^{3}\times \rho _{\text{ in FLRW}},
\end{equation}%
where $M_{\text{MK}}$ is just the same luminous lens mass $M$ appearing in
the metric (9).

We shall take the WMAP estimate of the observed density of the universe that
is the critical density $\rho _{\text{c}}$ implying that the universe is
spatially flat. \ Thus, for our computation, we take $\rho _{\text{ in FLRW}%
}=\rho _{\text{c}}=\frac{3H_{0}^{2}}{8\pi }=9.47\times 10^{-30}$ gm.cm$%
^{-3}=7.03\times 10^{-58}$ cm$^{-2}$ [80]. Normalizing the scale factor to $%
a(t_{0})=1$ and dropping suffixes, the above prescription translates to 
\begin{equation}
r_{b}=\left( \frac{3M}{4\pi \rho _{\text{c}}}\right) ^{1/3},
\end{equation}%
where the luminous mass $M$ is often expressed in units of sun's mass $%
M_{\odot }=1.989\times 10^{33}$ gm $=1.48\times 10^{5}$cm. Eq.(25) provides
a vacuole boundary radius $r_{b}$, where the spacetime transitions from a MK
spacetime to an FLRW background. Evidently, by the prescription (50), $r_{b}$
depends explicitly only on the observable cosmological Hubble parameter $%
H_{0}$ and on the galactic parameter $M$, none of which depends on the
constant $\gamma $. So, for consistency, the $\gamma $ terms in Eq.(47) [or,
in Eq.(48)] should be discarded. To verify what it entails, let us once
again note that, for typical galactic values from rotation curve fit [65] 
\begin{equation}
M\approx 10^{16}\text{ cm, }R\approx 10^{22}\text{ cm, }\gamma \approx
10^{-30}\text{ cm, }
\end{equation}%
and for $r_{b}\approx 10^{24}$ cm coming from the prescription (50), the
first piece in Eq.(48) turns out to be $\left( \frac{R}{r_{b}}-\frac{2M}{R}%
\right) \approx 10^{-2}$, while the last piece in the square bracket is $%
\approx 10^{-8}$, hence can be easily ignored in comparison. Thus, to the
leading order, $\varphi _{b}\approx 10^{-2}$, which is consistent with the
small angle approximation. Therefore, without any loss of rigor in what
follows, we can take%
\begin{equation}
\varphi _{b}\simeq \frac{R}{r_{b}}-\frac{2M}{R},
\end{equation}%
or equivalently,%
\begin{equation}
r_{b}\simeq \left( \frac{\varphi _{b}}{R}+\frac{2M}{R^{2}}\right) ^{-1}.
\end{equation}

Returning to Eq.(46), we get: 
\begin{eqnarray}
\left\vert A_{b}\right\vert &\equiv &\left\vert A(r_{b},\varphi
_{b})\right\vert  \notag \\
&=&\left( r_{b}^{2}\right) \times \left[ \frac{3MR\gamma (\varphi _{b}-2\pi
)\varphi _{b}+4(R-2M\varphi _{b})}{4R^{2}}\right]  \notag \\
&=&\frac{r_{b}^{2}}{R}\left[ 1-\frac{2M\varphi _{b}}{R}+\frac{3M\pi \gamma
\varphi _{b}}{4}-\frac{3M\gamma \varphi _{b}^{2}}{2}\right] >0.
\end{eqnarray}%

This positivity is possible as the observed galactic data $M,R$ and the
small values of $\varphi _{b}$ from Eq.(52) render the quantity in the
square bracket positive.

Ishak \textit{et al.}[66] formula for the bending angle $\psi $ is%
\begin{equation}
\tan \psi =\frac{r_{b}\sqrt{B(r_{b})}}{\left\vert A_{b}\right\vert },
\end{equation}%
where 
\begin{equation}
\sqrt{B(r_{b})}\simeq 1-\frac{M}{r_{b}}+\frac{1}{2}\gamma r_{b}-\frac{1}{2}%
kr_{b}^{2},
\end{equation}%
since, for galactic values of $M$, rotation curve-fitted values of $\gamma $%
, $k$ and the values of $r_{b}$, the last three terms add to $\approx
10^{-6} $, which is too small compared to unity justifying that the higher
power expansion terms in $\sqrt{B(r_{b})}$ be ignored. The main thing to
note is that $k$ appears in the bending only through $\sqrt{B(r_{b})}$.

For small $\psi $, $\tan \psi \simeq \psi $, and for small entry angle, $%
\tan \varphi _{b}\simeq \varphi _{b}$, so that the one-way deflection $%
\epsilon $ for nonzero $\varphi _{b}$ is, by definition [70]%
\begin{equation}
\epsilon =\tan (\psi -\varphi _{b})\simeq \psi -\varphi _{b},
\end{equation}%
where%
\begin{equation}
\psi =\frac{2R^{2}\left[ 2M+r_{b}\left( kr_{b}^{2}-\gamma r_{b}-2\right) %
\right] }{r_{b}^{2}\left[ 8M\varphi _{b}+R\left\{ 3M\gamma \varphi
_{b}\left( 2\varphi _{b}-\pi \right) -4\right\} \right] }.
\end{equation}%

Eq.(57) is the exact one-way expression but is rather unilluminating, so we
shall expand it in the first powers to see what it yields. Expanding Eq.(57)
in the first power of $\gamma $, we get, with $\psi \equiv \epsilon
_{0}+\epsilon _{1}$, 
\begin{eqnarray}
\epsilon &=&\left( \epsilon _{0}-\varphi _{b}\right) +\epsilon _{1}=\left[ 
\frac{2R^{2}\left( 2M-2Rr_{b}+kRr_{b}^{3}\right) }{r_{b}^{2}\left( 8M\varphi
_{b}-4R\right) }-\varphi _{b}\right] +\epsilon _{1}, \\
\epsilon _{1} &\equiv &\frac{2\gamma R^{2}}{r_{b}^{2}}\left[ \frac{r_{b}^{2}%
}{4R-8M\varphi _{b}}+\frac{3MR\left( 2M-2r_{b}+kRr_{b}^{3}\right) \left( \pi
-2\varphi _{b}\right) \varphi _{b}}{16\left( R-2M\varphi _{b}\right) ^{2}}%
\right] .
\end{eqnarray}%

Expanding $\epsilon _{1}$ in the first power of $M$, we get%
\begin{eqnarray}
\epsilon _{1} &\equiv &\frac{\gamma R}{2}+\frac{M}{8r_{b}}[8\gamma
r_{b}\varphi _{b}+3\pi kR\gamma r_{b}^{2}\varphi _{b}  \notag \\
&&+12R\gamma \varphi _{b}^{2}-6kR\gamma r_{b}^{2}\varphi _{b}^{2}-6\pi
R\gamma \varphi _{b}]+O(M^{2}) \\
&\simeq &\frac{\gamma R}{2},
\end{eqnarray}%
since, for typical galactic values of $M,R$, $\gamma $, $k$, $r_{b}$ and $%
\varphi _{b}$, it follows that $\frac{\gamma R}{2}\approx 10^{-7}$, while
the second term in Eq.(61) leads to a value $\approx 10^{-16}$ and so
ignored here by comparison. Expressing $r_{b}$ in terms of $\varphi _{b}$
using Eq.(52) in the square bracket of the right hand side of Eq.(59), we get

\begin{equation}
\epsilon _{0}-\varphi _{b}\equiv \frac{2\left( \frac{2M}{R}+\varphi
_{b}\right) ^{2}\left[ 2M-\frac{2R^{2}}{2M+R\varphi _{b}}+\frac{kR^{6}}{%
(2M+R\varphi _{b})^{3}}\right] }{8M\varphi _{b}-4R}-\varphi _{b}.
\end{equation}%

Expanding it in the first power of $M$, whence $\varphi _{b}$ on the right
hand side cancels out, we find%
\begin{equation}
\epsilon _{0}-\varphi _{b}=\frac{2M}{R}-\frac{kR^{2}}{2\varphi _{b}}+M\left( 
\frac{\varphi _{b}^{2}}{R}+\frac{kR}{\varphi _{b}^{2}}-kR\right) +O(M^{2}).
\end{equation}%

For typical galactic values mentioned above and for $k\approx 10^{-54}$ cm$%
^{-2}$, the relative strength of the terms in Eq.(64) are as follows:

\begin{equation}
\frac{2M}{R}\approx 10^{-6}\text{, \ }-\frac{kR^{2}}{2\varphi _{b}}\approx
-10^{-7}\text{, }M\left( \frac{\varphi _{b}^{2}}{R}+\frac{kR}{\varphi
_{b}^{2}}-kR\right) \approx 10^{-11}\text{. }
\end{equation}%

Hence, we can ignore the third term on the right hand side of Eq.(64).
Collecting the leading order terms from Eqs.(61) and (65), the result is 
\begin{equation}
\delta =2\epsilon =2\left[ \left( \epsilon _{0}-\varphi _{b}\right)
+\epsilon _{1}\right] =\frac{4M}{R}+\gamma R-\frac{kR^{2}}{\varphi _{b}}.
\end{equation}

In the limit, $\gamma =0$, one recovers the known bending expression for the
SdS metric [66]. The last cosmological term is expectedly repulsive and
looks familiar if one formally chooses $k=\Lambda /3$ so that, using
Eq.(52), one finds%
\begin{equation}
-\frac{kR^{2}}{\varphi _{b}}\simeq -\frac{\Lambda Rr_{b}}{3},
\end{equation}%
which is exactly the contribution obtained by Ishak \textit{et al.} [69].
Further notice that the term $+\gamma R$ in Eq.(66) enhances the
Schwarzschild bending $\frac{4M}{R}$, contrary to previous results, which is
what we wanted to prove.

Expressing $R$ in terms of the impact parameter $b$, and using Eq.(43), we
get\footnote{%
The impact parameter $b$ for the metric (9) is defined by [113]: $b=\left( 
\frac{1}{u_{\text{max}}}\right) \left[ \frac{1}{B(u_{\text{max}})}\right]
^{1/2}$, where $u_{\text{max}}=1/r_{\text{min}}=1/r_{0}$. Hence, $b=r_{0}%
\left[ \frac{1}{B(r_{0})}\right] ^{1/2}\simeq r_{0}+M-\frac{\gamma r_{0}^{2}%
}{2}+\frac{\Lambda r_{0}^{3}}{6}$. Ignoring $M$ compared to $r_{0}$ (as $%
M/r_{0}\approx 10^{-6}$) and inverting, one obtains $\frac{1}{b}\simeq \frac{%
1}{r_{0}}+\frac{\gamma }{2}-\frac{\Lambda r_{0}}{6}$. Solving for $\frac{1}{%
r_{0}}$, we get $\frac{1}{r_{0}}=\frac{\left( 6-3b\gamma \right) +\sqrt{%
\left( (6-3b\gamma \right) ^{2}+24b^{2}\Lambda }}{12b}$. Since $3b\gamma \ll
6$, we can write $\frac{1}{r_{0}}\simeq \frac{6+\sqrt{36+24b^{2}\Lambda }}{%
12b}=\frac{1}{b}+\frac{\Lambda b}{6}+O(b^{2})$, which is just our Eq.(68).
It also follows by expanding $R\simeq r_{0}=\left( \frac{1}{b}+\frac{\Lambda
b}{6}\right) ^{-1}$ in powers of $b$ that $R=b+O(b^{2})$.} 
\begin{equation}
\frac{1}{R}\simeq \frac{1}{r_{0}}\simeq \frac{1}{b}+\frac{\Lambda b}{6},
\end{equation}%
so we obtain, to leading order, from Eq.(66) 
\begin{eqnarray}
\delta &=&\frac{4M}{b}+\frac{2M\Lambda b}{3}+\gamma b-\frac{\Lambda br_{b}}{3%
} \\
&\equiv &t_{\text{Sch}}+t_{\text{Sereno}}+t_{\gamma }+t_{\Lambda }.
\end{eqnarray}%

The term $\frac{2M\Lambda b}{3}$ has been obtained by Sereno [111] and he
called it a local coupling term. There is also another term proportional to $%
M^{2}$ adding to $\delta $, viz., $\frac{15M^{2}\gamma }{b}$ derived
previously [112] but not shown here. However, for galactic lenses their
values are too minute to be of any interest, e.g., $t_{\text{Sereno}}\approx
10^{-15}$, as well as the still smaller other term, $\frac{15M^{2}\gamma }{b}%
\approx 10^{-20}$, so we ignore them here. But the remaining terms are
nearly of comparable magnitude, so we preserve them. Restoring $k$, we
therefore have: 
\begin{equation}
\delta =\frac{4M}{b}+\gamma b-kbr_{b}\equiv t_{\text{Sch}}+t_{\gamma }+t_{k}%
\text{. }
\end{equation}%
\textit{This is our final expression for light bending in the MK solution to
be used in our analysis of the galactic mass decomposition. }

\section{Algorithm for mass decomposition}

The luminous mass $M$ in the metric (9) will hereafter be denoted by $%
M_{\ast }$ for more notational clarity. Thus, the MK light deflection $%
\delta $, Eq.(71), can be rewritten as 
\begin{eqnarray}
\delta &=&\frac{4M_{\ast }}{b}+\widetilde{\gamma }b, \\
&=&t_{\text{Sch}}+t_{\gamma }+t_{k}
\end{eqnarray}%
where%
\begin{equation*}
t_{\text{Sch}}=\frac{4M_{\ast }}{b}\text{, }t_{\gamma }=\gamma b=(N^{\ast
}\gamma ^{\ast }+\gamma _{0})b\text{, }t_{k}=-kbr_{b}
\end{equation*}%
\begin{equation}
M_{\ast }=N^{\ast }\beta ^{\ast }
\end{equation}%
\begin{equation}
\gamma =N^{\ast }\gamma ^{\ast }+\gamma _{0}
\end{equation}%
\begin{equation}
\widetilde{\gamma }\equiv \gamma -kr_{b},\text{ }
\end{equation}%
where the solar mass is $\beta ^{\ast }=1.48\times 10^{5}$ cm, and the
universal constants in the MK solution are: $\gamma ^{\ast }=5.42\times
10^{-41}$ cm$^{-1}$, $\gamma _{0}=3.06\times 10^{-30}$ cm$^{-1}$, $%
k=9.54\times 10^{-54}$ cm$^{-2}$. Table I below will show the values of
different contributions to $\delta $ in the case of 57 lens galaxies as well
as the corresponding values of $b$ and $r_{b}$.

On the other hand, the Schwarzschild deflection $\alpha $ in Einstein theory
is 
\begin{equation}
\alpha =\frac{4M_{\text{tot}}^{\text{lens}}}{b},
\end{equation}%
where $M_{\text{tot}}^{\text{lens}}$ is the total projected lens mass
enclosed within the Einstein radius $R_{\text{Ein}}$ defined by $R_{\text{Ein%
}}=$ $d_{\text{ol}}\theta _{\text{Ein}}$, which is nothing but the impact
parameter $b$. We shall imbed the deflection expressions into the lens
equation [116], which is 
\begin{equation}
\theta d_{\text{os}}=\beta d_{\text{os}}+\alpha d_{\text{ls}},
\end{equation}%
where $d_{\text{os}}$, $d_{\text{ls}}$, $d_{\text{ol}}$ are the angular
diameter distances between observer-source, lens-source and observer-lens.
The Einstein angle $\theta =\theta _{\text{Ein}}$ is defined by the case
when the source, lens and observer stay in a line, that is, when $\beta =0$.
Thus%
\begin{eqnarray}
\theta &=&\alpha \left( \frac{d_{\text{ls}}}{d_{\text{os}}}\right) =\frac{%
4M_{\text{tot}}^{\text{lens}}}{b}\left( \frac{d_{\text{ls}}}{d_{\text{os}}}%
\right) =\frac{4M_{\text{tot}}^{\text{lens}}}{\theta d_{\text{ol}}}\left( 
\frac{d_{\text{ls}}}{d_{\text{os}}}\right) \\
&\Rightarrow &\theta =\theta _{\text{Ein}}=\sqrt{\frac{4M_{\text{tot}}^{%
\text{lens}}}{D}}, \\
D &\equiv &\frac{d_{\text{ol}}d_{\text{os}}}{d_{\text{ls}}}.
\end{eqnarray}%

Note that $\theta _{\text{Ein}}$ is caused by the total mass (luminous $%
M_{\ast }$ + dark) enclosd within the Einstein radius $b$.

The Weyl angle, for the \textit{same} impact parameter $b=\theta d_{\text{ol}%
}$ is%
\begin{eqnarray}
\theta &=&\delta \left( \frac{d_{\text{ls}}}{d_{\text{os}}}\right) =\left[ 
\frac{4M_{\ast }}{\theta d_{\text{ol}}}+\widetilde{\gamma }(\theta d_{\text{%
ol}})\right] \left( \frac{d_{\text{ls}}}{d_{\text{os}}}\right) \\
&\Rightarrow &\theta =\theta _{\text{Weyl}}=\sqrt{\frac{4M_{\ast }}{D-%
\widetilde{\gamma }d_{\text{ol}}^{2}}}.
\end{eqnarray}%

Note that $\theta _{\text{Weyl}}$ is caused by the luminous mass $M_{\ast }$
plus cosmology induced potentials within the same radius $b$ (i.e., same
lensing geometry, same impact parameter). Following Edery and Paranjape
[72], and later works [48,49], we use the input $\theta _{\text{Ein}}=$ $%
\theta _{\text{Weyl}}$, which yields%
\begin{equation}
\widetilde{\gamma }=\frac{d_{\text{os}}}{d_{\text{ls}}d_{\text{ol}}}\left( 1-%
\frac{M_{\ast }}{M_{\text{tot}}^{\text{lens}}}\right) .
\end{equation}%

Using Eq.(50) for $r_{b}$ in Eq.(76), Eq.(84) can be explicity written as 
\begin{equation}
N^{\ast }\gamma ^{\ast }+\gamma _{0}=\frac{d_{\text{os}}}{d_{\text{ls}}d_{%
\text{ol}}}\left( 1-\frac{N^{\ast }\beta ^{\ast }}{M_{\text{tot}}^{\text{lens%
}}}\right) +\frac{k\left( \frac{3}{\pi }\right) ^{1/3}}{2^{5/3}}\left( \frac{%
N^{\ast }\beta ^{\ast }}{\rho _{\text{c}}}\right) ^{1/3},
\end{equation}%
where [112]%
\begin{equation*}
\rho _{\text{c}}=\frac{3H_{0}^{2}}{8\pi }=9.47\times 10^{-30}\text{ gm.cm}%
^{-3}=7.03\times 10^{-58}\text{ cm}^{-2}\text{.}
\end{equation*}%
\textit{Eq.(85) is a cubic equation in }$N^{\ast }$\textit{\ and is central
to our mass decomposition scheme.}

Our algorithm is the following: In the above Eq.(85), the galaxy independent
universal MK constants ($\gamma _{0}$, $\gamma ^{\ast }$, $k$) are known
[62,65], the distances $d_{\text{ls}}$, $d_{\text{ol}}$, $d_{\text{os}}$ ($%
\equiv d_{\text{ol}}+d_{\text{ls}}$) and the total mass $M_{\text{tot}}^{%
\text{lens}}$ are provided by the observed SLACS data for each individual
galaxy [73,119]. Plugging in these values in Eq.(85), we first have to find
the numerical value of $N^{\ast }$ specific to each galaxy. The resultant
cubic equation in $N^{\ast }$ fortunately yields only one positive root,
which then enables us to find the value of the luminous component $M_{\ast
}= $ $N^{\ast }\beta ^{\ast }$. We henceforth call it $M_{\ast }^{\text{MK}}$
to distinguish it from the luminous mass values obtained from other
independent simulations. Subtracting $M_{\ast }^{\text{MK}}$ from the
observed total mass of the lens $M_{\text{tot}}^{\text{lens}}$, we obtain
the dark matter component $M_{\text{dm}}$ as well as the mass ratios $%
f_{\ast }^{\text{MK}}=(M_{\ast }^{\text{MK}}/M_{\text{tot}}^{\text{lens}%
})|_{\leq R_{\text{Ein}}}$.

As to the existing mass ratios in the literature, we note that simulations
depending on different stellar-population models and IMFs have thrown up
rather widely different values with error bars. This can be seen from the
work of Grillo \textit{et al.} [73], who fitted the lens spectral energy
distributions (SEDs) with a three-parameter grid of Bruzual \& Charlot's
(indexed BC) [116] and Maraston's (indexed M) [117] composite
stellar-population models, computed by adopting solar metallicity and
various initial mass functions (IMFs). They obtained the mass decomposition
within the Einstein radius and found that a Salpeter IMF (indexed Sal) [118]
was better suited than either a Chabrier (Cha) [119] or Kroupa (Kro) IMF
[120] for describing the considered subsample of $57$ lenses. It was
concluded that in all the models, the observed total mass $M_{\text{tot}}^{%
\text{lens}}$ is linearly proportional to the estimated luminous mass of the
lenses denoted by $M_{\ast }^{\text{Sal,BC}}$, $M_{\ast }^{\text{Sal,M}}$, $%
M_{\ast }^{\text{Cha,BC}}$ and $M_{\ast }^{\text{Kro,M}}$. However, the dark
matter component was found to be considerably higher for the two models
[(Cha,BC), (Kro,M)] than for the models [(Sal,BC), (Sal,M)].

Our results on mass decomposition are shown in Table II below. We find that
the mass ratios generated from the better suited model (Sal,BC), viz., $%
f_{\ast }^{\text{Grillo \textit{et al.}}}=(M_{\ast }^{\text{Sal,BC}}/M_{%
\text{tot}}^{\text{lens}})|_{\leq R_{\text{Ein}}}$ [73] come closer to our
computed ratios than those from other models. For the $57$ galaxies, the
values of $f_{\ast }^{\text{Grillo \textit{et al.}}}$ range from $0.43$ to $%
1.21$, which can be compared with the ratios $f_{\ast }^{\text{MK}}$ that
are seen to take on values very near unity ( $f_{\ast }^{\text{MK}}\sim 0.98$
on the average) for all these galaxies (Table II). These values show that
the matter content within the Einstein radius of the lens is not dominated
by dark matter. However, it is evident from the Table II that $M_{\text{tot}%
}^{\text{lens}}$ is linearly proportional to $M_{\ast }^{\text{MK}}$ across
the entire subsample, which is in complete agreement with the common
prediction of average linearity by other models (Figs.1 \& 2) illustrated by
a line parallel to the one-to-one line defined by $M_{\ast }/M_{\text{tot}}^{%
\text{lens}}=1$, where $M_{\ast }$ could be $M_{\ast }^{\text{MK}}$, $%
M_{\ast }^{\text{Sal,BC}}$ or $M_{\ast }^{\text{Cha,BC}}$ etc.\footnote{%
It should be mentioned that the \textit{exact} equality $M_{\ast }/M_{\text{%
tot}}^{\text{lens}}=1$ is inconsistent with SDSS data. For the subsample
under study, photometric and spectroscopic data are available. By using the
SDSS multicolor photometry and lens modeling, Grillo\textit{\ et al.} [73]
studied the luminous and dark matter composition in the sample. It is
possible for the data to be consistent with $M_{\ast }/M_{\text{tot}}^{\text{%
lens}}\approx 1$ allowing for dark matter however little, as is the case
with a few galaxies (Fig.1a) lying almost on the one-to-one line.} Fig.1
compares the average linear profile of the model $f_{\ast }^{\text{Grillo 
\textit{et al.}}}$ with the more exact linearity of $f_{\ast }^{\text{MK}}$
denoted by dots sitting just above the one-to-one line$.$

In detail, we find that $17$ galaxies show mass ratios that fall within the
(Sal,BC) projected error bars shown in [73]. Interestingly, out of these $17$%
, we find that $5$ galaxies exhibit very little dark matter, $4$ of which
are supported by the (Sal,BC) model, which yield $f_{\ast }^{\text{MK}%
}=f_{\ast }^{\text{Grillo \textit{et al.}}}\approx 1$, as marked the
coincident points in Fig.1a. The remaining one galaxy $(J0959+0410)$ falls
outside the one-to-one line by the margin of an additive factor $0.28$. The
ratios for the $22$ galaxies fall marginally outside the error bar, while
for the remaining $18$ galaxies the ratios fall outside by a maximum margin $%
0.52.$ Fig.2 shows linearity profiles for the two other models [(Cha,BC),
(Kro,M)] that throw up higher amounts of dark matter within the Einstein
radius (the higher is the average line over the one-to-one line, the more is
the dark matter). The profiles can be compared by noting that the linear fit
by Grillo [123, his \textit{Eq.(7)}] is

\begin{equation}
\text{Log}_{10}[M_{\text{tot}}^{\text{lens}}(R_{\text{Ein}})]=-0.58+1.09\ast 
\text{Log}_{10}[M_{\ast }(R_{\text{Ein}})].
\end{equation}%

There is an average difference of about $0.36$ with our line, which fits to

\begin{equation}
\text{Log}_{10}[M_{\text{tot}}^{\text{lens}}(R_{\text{Ein}})]=-0.94+1.09\ast 
\text{Log}_{10}[M_{\ast }(R_{\text{Ein}})].
\end{equation}

In view of the above, our prediction of non-dominance of dark matter within
the Einstein radius suggests that the present analytic approach is more akin
to models [(Sal,BC), (Sal,M)] that provide relatively low dark matter inside
the Einstein radius though all of their predicted ratios do not exactly
coincide with, but not stray far away from either, those from our approach.
Given the model-dependent varying mass decompositions in the literature and
lack of any direct experimental support yet, we can regard our
model-independent analytic approach as an alternative scheme for deriving
mass decompositions.

The mean density $\left\langle \rho \right\rangle _{\text{av}}^{\text{MK}}$
is obtained by averaging the dark matter mass $M_{\text{dm}}^{\text{MK}}=M_{%
\text{tot}}^{\text{lens}}(\leq R_{\text{Ein}})-M_{\ast }^{\text{MK}}(\leq R_{%
\text{Ein}})$ over the Einstein sphere of radius $R_{\text{Ein}}$ centered
at the galactic origin: 
\begin{equation}
\left\langle \rho \right\rangle _{\text{av}}^{\text{MK}}=\frac{3M_{\text{dm}%
}^{\text{MK}}}{4\pi b^{3}}.
\end{equation}%

Table I includes the values of $\left\langle \rho \right\rangle _{\text{av}%
}^{\text{MK}}$ together with the values of the impact parameter and the
relevant deflection components. Table II gives the estimates of stellar and
dark matter masses together with $f_{\ast }^{\text{MK}}$. Data for distances 
$d_{\text{os}}$, $d_{\text{ol}}$ and $M_{\text{tot}}^{\text{lens}}$ are
taken from [117]. The conversions used are: $1$ arcsec $=(1/206265)$ rad, $1$
Mpc $=3.085\times 10^{24}$ cm, $1$ cm$^{-2}=1.98\times 10^{59}M_{\odot }($kpc%
$)^{-3}$.

\begin{center}
\textbf{Table I. Different contributions to Schwarzschild light bending $t_{\text{Sch%
}}$ with the associated impact parameter $b$. The last two columns show that
the positive contribution $t_{\gamma }$ is overtaken by the negative
contribution $t_{k}$ in all the cases. However, their combined effect is one
or two orders of magnitude less than the contribution $t_{\text{Sch}}$. The
overall bending is thus positive and towards the galactic center. The
average dark matter density $\left\langle \rho \right\rangle _{\text{av}}^{%
\text{MK}}$ over the Einstein sphere of radius $b$ is seen to be $\sim
10^{6}M_{\odot }$(kpc)$^{-3}$.}
\begin{longtable}{|l|l|l|l|l|l|}
\hline
Galaxy & $\left\langle \rho \right\rangle _{\text{av}}^{\text{MK}}$ & $b$ & $%
t_{\text{Sch}}$ & $t_{\gamma }$ & $t_{k}$ \\ \hline
- & $\times 10^{6}M_{\odot }$(kpc)$^{-3}$ & (kpc) & $\times 10^{-6}$ & $%
\times 10^{-7}$ & $\times 10^{-7}$ \\ \hline
J0008-0004 & $5.77$ & $6.59$ & $9.98$ & $4.40$ & $-5.05$ \\ \hline
J0029-0055 & $0.93$ & $3.49$ & $6.58$ & $1.03$ & $-1.87$ \\ \hline
J0037-0942 & $5.51$ & $4.97$ & $11.09$ & $2.85$ & $-3.57$ \\ \hline
J0044+0113 & $0.17$ & $1.71$ & $10.11$ & $0.42$ & $-0.83$ \\ \hline
J0109+1500 & $2.15$ & $3.03$ & $8.21$ & $0.94$ & $-1.66$ \\ \hline
J0157-0056 & $6.70$ & $4.88$ & $10.07$ & $2.55$ & $-3.38$ \\ \hline
J0216-0813 & $17.84$ & $5.53$ & $16.54$ & $4.93$ & $-4.73$ \\ \hline
J0252+0039 & $2.48$ & $4.41$ & $7.78$ & $1.73$ & $-2.70$ \\ \hline
J0330-0020 & $3.88$ & $5.44$ & $8.72$ & $2.76$ & $-3.72$ \\ \hline
J0405-0455 & $0.00$ & $1.14$ & $5.05$ & $0.16$ & $-0.38$ \\ \hline
J0728+3835 & $3.27$ & $4.22$ & $9.04$ & $1.80$ & $-2.68$ \\ \hline
J0737+3216 & $8.16$ & $4.67$ & $11.76$ & $2.68$ & $-3.36$ \\ \hline
J0822+2652 & $5.02$ & $4.45$ & $10.26$ & $2.19$ & $-3.00$ \\ \hline
J0903+4116 & $7.93$ & $7.23$ & $11.59$ & $5.97$ & $-6.02$ \\ \hline
J0912+0029 & $12.70$ & $4.58$ & $16.51$ & $3.46$ & $-3.67$ \\ \hline
J0935-0003 & $25.07$ & $4.27$ & $18.04$ & $3.27$ & $-3.44$ \\ \hline
J0936+0913 & $2.15$ & $3.45$ & $8.31$ & $1.19$ & $-1.99$ \\ \hline
J0946+1006 & $5.94$ & $4.93$ & $11.15$ & $2.83$ & $-3.54$ \\ \hline
J0956+5100 & $10.26$ & $5.04$ & $13.86$ & $3.55$ & $-3.93$ \\ \hline
J0959+4416 & $3.21$ & $3.61$ & $9.01$ & $1.36$ & $-2.17$ \\ \hline
J0959+0410 & $0.00$ & $2.23$ & $6.87$ & $0.51$ & $-1.04$ \\ \hline
J1016+3859 & $2.57$ & $3.13$ & $9.18$ & $1.07$ & $-1.80$ \\ \hline
J1020+1122 & $8.80$ & $5.12$ & $12.56$ & $3.35$ & $-3.88$ \\ \hline
J1023+4230 & $3.96$ & $4.48$ & $9.77$ & $2.13$ & $-2.98$ \\ \hline
J1029+0420 & $0.00$ & $1.93$ & $5.97$ & $0.37$ & $-0.82$ \\ \hline
J1100+5329 & $8.39$ & $7.03$ & $12.49$ & $6.04$ & $-5.93$ \\ \hline
J1106+5228 & $0.08$ & $2.18$ & $7.89$ & $0.53$ & $-1.06$ \\ \hline
J1112+0826 & $9.89$ & $6.21$ & $13.59$ & $5.15$ & $-5.16$ \\ \hline
J1134+6027 & $1.71$ & $2.92$ & $8.52$ & $0.91$ & $-1.60$ \\ \hline
J1142+1001 & $3.36$ & $3.50$ & $9.27$ & $1.32$ & $-2.11$ \\ \hline
J1143-0144 & $3.80$ & $3.26$ & $11.14$ & $1.34$ & $-2.03$ \\ \hline
J1153+4612 & $0.65$ & $3.18$ & $6.62$ & $0.88$ & $-1.65$ \\ \hline
J1204+0358 & $2.55$ & $3.68$ & $8.82$ & $1.39$ & $-2.21$ \\ \hline
J1205+4910 & $5.86$ & $4.25$ & $11.18$ & $2.16$ & $-2.91$ \\ \hline
J1213+6708 & $1.72$ & $3.13$ & $8.54$ & $1.02$ & $-1.76$ \\ \hline
J1218+0830 & $2.25$ & $3.47$ & $8.82$ & $1.25$ & $-2.04$ \\ \hline
J1250+0523 & $2.62$ & $4.18$ & $8.22$ & $1.64$ & $-2.56$ \\ \hline
J1402+6321 & $7.05$ & $4.54$ & $12.13$ & $2.61$ & $-3.26$ \\ \hline
J1403+0006 & $0.53$ & $2.62$ & $7.32$ & $0.68$ & $-1.32$ \\ \hline
J1416+5136 & $7.03$ & $6.08$ & $11.45$ & $4.27$ & $-4.74$ \\ \hline
J1420+6019 & $0.00$ & $1.26$ & $6.08$ & $0.20$ & $-0.46$ \\ \hline
J1430+4105 & $1.31$ & $6.53$ & $15.41$ & $6.34$ & $-5.77$ \\ \hline
J1436-0000 & $3.97$ & $4.81$ & $9.09$ & $2.29$ & $-3.20$ \\ \hline
J1443+0304 & $0.00$ & $1.92$ & $5.98$ & $0.37$ & $-0.81$ \\ \hline
J1451-0239 & $0.00$ & $2.33$ & $6.58$ & $0.53$ & $-1.09$ \\ \hline
J1525+3327 & $11.08$ & $6.56$ & $13.64$ & $5.74$ & $-5.58$ \\ \hline
J1531-0105 & $4.84$ & $4.71$ & $10.89$ & $2.55$ & $-3.31$ \\ \hline
J1538+5817 & $0.07$ & $2.51$ & $6.87$ & $0.61$ & $-1.22$ \\ \hline
J1621+3931 & $6.09$ & $4.97$ & $11.07$ & $2.85$ & $-3.57$ \\ \hline
J1627-0053 & $4.88$ & $4.18$ & $10.47$ & $1.99$ & $-2.78$ \\ \hline
J1630+4520 & $8.66$ & $6.91$ & $13.25$ & $6.18$ & $-5.92$ \\ \hline
J1636+4707 & $2.95$ & $3.98$ & $8.64$ & $1.56$ & $-2.43$ \\ \hline
J2238-0754 & $1.44$ & $3.07$ & $8.09$ & $0.95$ & $-1.69$ \\ \hline
J2300+0022 & $8.06$ & $4.54$ & $12.54$ & $2.68$ & $-3.30$ \\ \hline
J2303+1422 & $5.78$ & $4.35$ & $11.81$ & $2.36$ & $-3.05$ \\ \hline
J2321-0939 & $1.31$ & $2.47$ & $9.31$ & $0.72$ & $-1.32$ \\ \hline
J2341+0000 & $3.48$ & $4.48$ & $9.35$ & $2.06$ & $-2.94$ \\ \hline
\end{longtable}

\bigskip

\textbf{Table II. Lens mass decomposition by the algorithm Eq.(85) and the vacuole
radius $r_{b}$ for individual galaxies.}

\begin{longtable}{|l|l|l|l|l|l|l|}
\hline
Galaxy & $M_{\ast }^{\text{MK}}$ & $M_{\text{dm}}^{\text{MK}}$ & $M_{\text{%
tot}}^{\text{lens}}$ & $r_{b}$ & $f_{\ast \text{Sal,BC}}^{\text{Grillo}}$ & $%
f_{\ast }^{\text{MK}}$ \\ \hline
- & ($\times 10^{11}M_{\odot }$) & ($\times 10^{8}M_{\odot }$) & ($\times
10^{11}M_{\odot }$) & (kpc) &  &  \\ \hline
J0008-0004 & $3.43$ & $69.12$ & $3.50$ & $842.98$ & $0.54_{-0.33}^{+0.10}$ & 
$0.98$ \\ \hline
J0029-0055 & $1.19$ & $1.65$ & $1.20$ & $590.01$ & $0.76_{-0.23}^{+0.35}$ & $%
0.99$ \\ \hline
J0037-0942 & $2.87$ & $28.34$ & $2.90$ & $791.76$ & $0.74_{-0.28}^{+0.17}$ & 
$0.99$ \\ \hline
J0044+0113 & $0.89$ & $0.03$ & $0.90$ & $536.05$ & $0.55_{-0.15}^{+0.27}$ & $%
0.99$ \\ \hline
J0109+1500 & $1.29$ & $2.51$ & $1.3$ & $605.96$ & $1.08_{-0.22}^{+0.29}$ & $%
0.99$ \\ \hline
J0157-0056 & $2.56$ & $32.85$ & $2.6$ & $763.46$ & $1.21_{-0.52}^{+0.20}$ & $%
0.98$ \\ \hline
J0216-0813 & $4.77$ & $126.99$ & $4.9$ & $943.08$ & $0.71_{-0.28}^{+0.19}$ & 
$0.97$ \\ \hline
J0252+0039 & $1.79$ & $8.97$ & $1.8$ & $675.39$ & $0.52_{-0.24}^{+0.07}$ & $%
0.99$ \\ \hline
J0330-0020 & $2.47$ & $26.32$ & $2.5$ & $753.54$ & $0.99_{-0.27}^{+0.11}$ & $%
0.99$ \\ \hline
J0405-0455 & $0.30$ & $0.00$ & $0.3$ & $371.68$ & $0.73_{-0.23}^{+0.43}$ & $%
1.00$ \\ \hline
J0728+3835 & $1.99$ & $10.33$ & $2.0$ & $699.53$ & $0.50_{-0.08}^{+0.25}$ & $%
0.99$ \\ \hline
J0737+3216 & $2.86$ & $34.99$ & $2.90$ & $791.76$ & $0.77_{-0.17}^{+0.09}$ & 
$0.98$ \\ \hline
J0822+2652 & $2.38$ & $18.59$ & $2.4$ & $743.36$ & $0.93_{-0.16}^{+0.10}$ & $%
0.99$ \\ \hline
J0903+4116 & $4.37$ & $126.14$ & $4.5$ & $916.64$ & $0.87_{-0.31}^{+0.09}$ & 
$0.97$ \\ \hline
J0912+0029 & $3.95$ & $51.46$ & $4$ & $881.35$ & $0.51_{-0.09}^{+0.11}$ & $%
0.98$ \\ \hline
J0935-0003 & $4.01$ & $82.12$ & $4.1$ & $888.64$ & $0.52_{-0.12}^{+0.09}$ & $%
0.98$ \\ \hline
J0936+0913 & $1.49$ & $3.72$ & $1.5$ & $635.56$ & $0.71_{-0.17}^{+0.21}$ & $%
0.99$ \\ \hline
J0946+1006 & $2.87$ & $30.02$ & $2.9$ & $791.76$ & $0.51_{-0.11}^{+0.06}$ & $%
0.99$ \\ \hline
J0956+5100 & $3.64$ & $55.33$ & $3.7$ & $858.74$ & $0.74_{-0.33}^{+0.11}$ & $%
0.98$ \\ \hline
J0959+4416 & $1.69$ & $6.33$ & $1.7$ & $662.64$ & $0.82_{-0.20}^{+0.19}$ & $%
0.99$ \\ \hline
J0959+0410 & $0.80$ & $0.00$ & $0.8$ & $515.42$ & $0.65_{-0.15}^{+0.13}$ & $%
1.00$ \\ \hline
J1016+3859 & $1.49$ & $3.31$ & $1.5$ & $635.56$ & $0.63_{-0.18}^{+0.16}$ & $%
0.99$ \\ \hline
J1020+1122 & $3.35$ & $49.56$ & $3.4$ & $834.88$ & $0.46_{-0.12}^{+0.19}$ & $%
0.98$ \\ \hline
J1023+4230 & $2.28$ & $15.01$ & $2.3$ & $732.89$ & $0.71_{-0.15}^{+0.09}$ & $%
0.99$ \\ \hline
J1029+0420 & $0.60$ & $.0.00$ & $0.6$ & $468.29$ & $0.81_{-0.27}^{+0.32}$ & $%
1.00$ \\ \hline
J1100+5329 & $4.57$ & $122.47$ & $4.7$ & $930.03$ & $0.49_{-0.15}^{+0.32}$ & 
$0.97$ \\ \hline
J1106+5228 & $0.89$ & $0.04$ & $0.9$ & $536.05$ & $1.01_{-0.29}^{+0.38}$ & $%
0.99$ \\ \hline
J1112+0826 & $4.40$ & $99.48$ & $4.5$ & $916.64$ & $0.70_{-0.10}^{+0.10}$ & $%
0.97$ \\ \hline
J1134+6027 & $1.29$ & $1.79$ & $1.3$ & $605.96$ & $0.65_{-0.25}^{+0.28}$ & $%
0.99$ \\ \hline
J1142+1001 & $1.69$ & $6.08$ & $1.7$ & $662.64$ & $0.59_{-0.13}^{+0.24}$ & $%
0.99$ \\ \hline
J1143-0144 & $1.89$ & $5.53$ & $1.9$ & $687.67$ & $0.46_{-0.06}^{+0.09}$ & $%
0.99$ \\ \hline
J1153+4612 & $1.09$ & $0.89$ & $1.1$ & $573.14$ & $0.51_{-0.07}^{+0.34}$ & $%
0.99$ \\ \hline
J1204+0358 & $1.69$ & $5.37$ & $1.7$ & $662.64$ & $0.43_{-0.12}^{+0.14}$ & $%
0.99$ \\ \hline
J1205+4910 & $2.48$ & $18.99$ & $2.5$ & $753.54$ & $0.61_{-0.19}^{+0.19}$ & $%
0.99$ \\ \hline
J1213+6708 & $1.39$ & $2.23$ & $1.4$ & $621.12$ & $0.71_{-0.18}^{+0.21}$ & $%
0.99$ \\ \hline
J1218+0830 & $1.59$ & $3.94$ & $1.6$ & $649.38$ & $0.64_{-0.14}^{+0.21}$ & $%
0.99$ \\ \hline
J1250+0523 & $1.79$ & $8.03$ & $1.8$ & $675.39$ & $1.04_{-0.33}^{+0.27}$ & $%
0.99$ \\ \hline
J1402+6321 & $2.87$ & $27.71$ & $2.9$ & $791.76$ & $0.73_{-0.12}^{+0.13}$ & $%
0.99$ \\ \hline
J1403+0006 & $0.99$ & $0.40$ & $1$ & $555.22$ & $0.84_{-0.30}^{+0.27}$ & $%
0.99$ \\ \hline
J1416+5136 & $3.63$ & $66.60$ & $3.7$ & $858.74$ & $0.75_{-0.17}^{+0.09}$ & $%
0.98$ \\ \hline
J1420+6019 & $0.40$ & $0.00$ & $0.4$ & $409.08$ & $1.01_{-0.36}^{+0.43}$ & $%
1.00$ \\ \hline
J1430+4105 & $5.24$ & $153.07$ & $5.4$ & $974.08$ & $0.38_{-0.13}^{+0.07}$ & 
$0.97$ \\ \hline
J1436-0000 & $2.28$ & $18.57$ & $2.3$ & $732.89$ & $0.77_{-0.30}^{+0.24}$ & $%
0.99$ \\ \hline
J1443+0304 & $0.60$ & $0.00$ & $0.6$ & $468.29$ & $1.00_{-0.44}^{+0.32}$ & $%
1.00$ \\ \hline
J1451-0239 & $0.80$ & $0.00$ & $0.8$ & $515.42$ & $0.97_{-0.48}^{+0.22}$ & $%
1.00$ \\ \hline
J1525+3327 & $4.66$ & $131.70$ & $4.8$ & $936.57$ & $0.68_{-0.21}^{+0.09}$ & 
$0.97$ \\ \hline
J1531-0105 & $2.68$ & $21.34$ & $2.7$ & $773.13$ & $0.70_{-0.14}^{+0.15}$ & $%
0.99$ \\ \hline
J1538+5817 & $0.89$ & $0.05$ & $0.9$ & $536.05$ & $0.84_{-0.19}^{+0.08}$ & $%
0.99$ \\ \hline
J1621+3931 & $2.87$ & $31.39$ & $2.9$ & $791.76$ & $0.75_{-0.25}^{+0.12}$ & $%
0.98$ \\ \hline
J1627-0053 & $2.28$ & $15.03$ & $2.3$ & $732.89$ & $0.61_{-0.13}^{+0.12}$ & $%
0.99$ \\ \hline
J1630+4520 & $4.78$ & $120.36$ & $4.9$ & $943.03$ & $0.69_{-0.10}^{+0.07}$ & 
$0.97$ \\ \hline
J1636+4707 & $1.79$ & $7.80$ & $1.8$ & $675.39$ & $0.59_{-0.12}^{+0.18}$ & $%
0.99$ \\ \hline
J2238-0754 & $1.29$ & $1.77$ & $1.3$ & $605.96$ & $0.64_{-0.25}^{+0.25}$ & $%
0.99$ \\ \hline
J2300+0022 & $2.97$ & $31.68$ & $3$ & $800.76$ & $0.60_{-0.11}^{+0.07}$ & $%
0.98$ \\ \hline
J2303+1422 & $2.68$ & $20.01$ & $2.7$ & $773.13$ & $0.63_{-0.09}^{+0.13}$ & $%
0.99$ \\ \hline
J2321-0939 & $1.19$ & $0.83$ & $1.2$ & $590.01$ & $0.90_{-0.18}^{+0.26}$ & $%
0.99$ \\ \hline
J2341+0000 & $2.18$ & $13.22$ & $2.2$ & $722.11$ & $0.73_{-0.28}^{+0.13}$ & $%
0.99$ \\ \hline
\end{longtable}
\end{center}

\begin{figure}[ht]
\begin{minipage}[h]{0.55\linewidth}
\center{\includegraphics[width=1\linewidth]{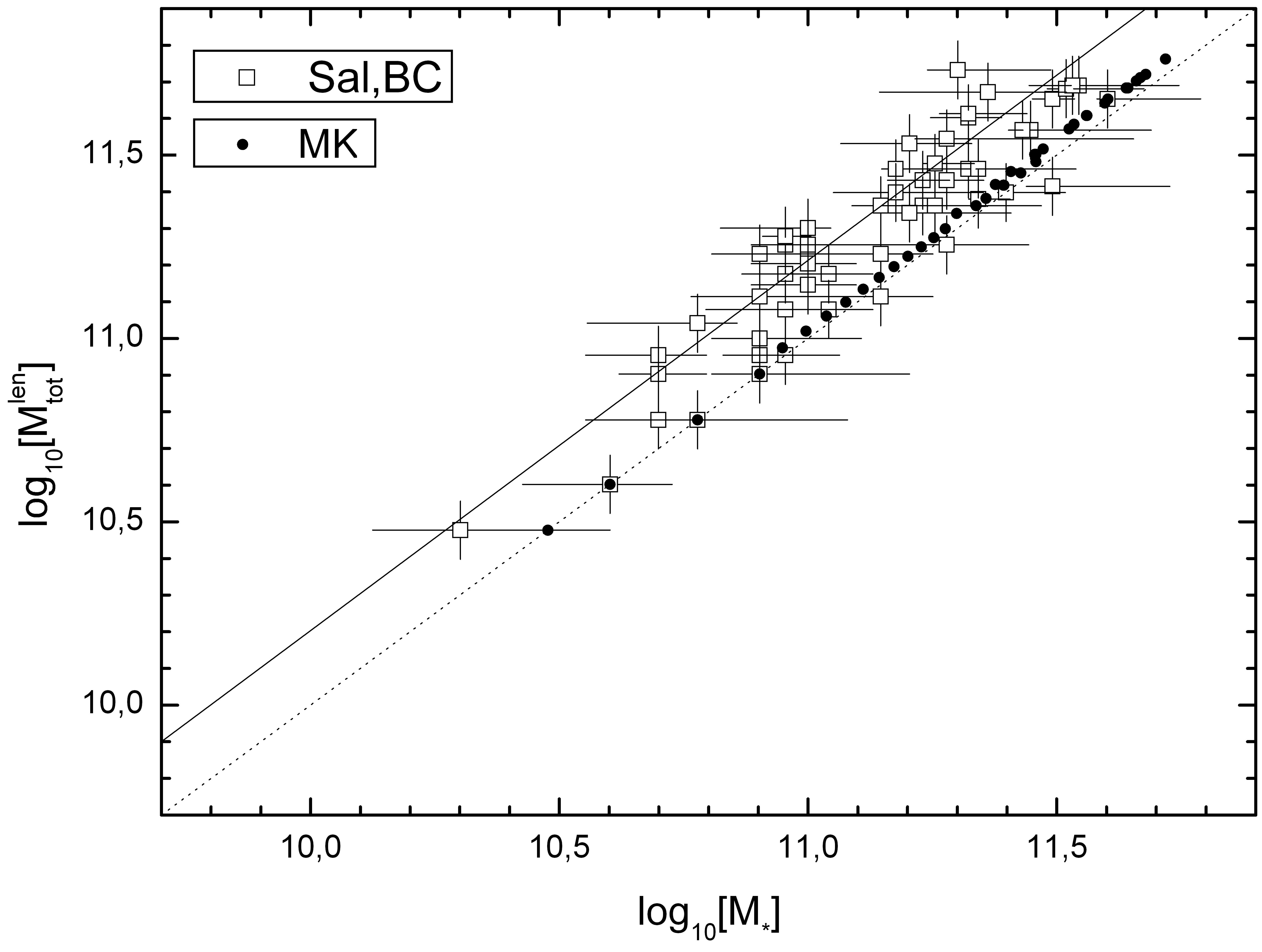}} a) \\
\end{minipage}
\hfill
\begin{minipage}[h]{0.55\linewidth}
\center{\includegraphics[width=1\linewidth]{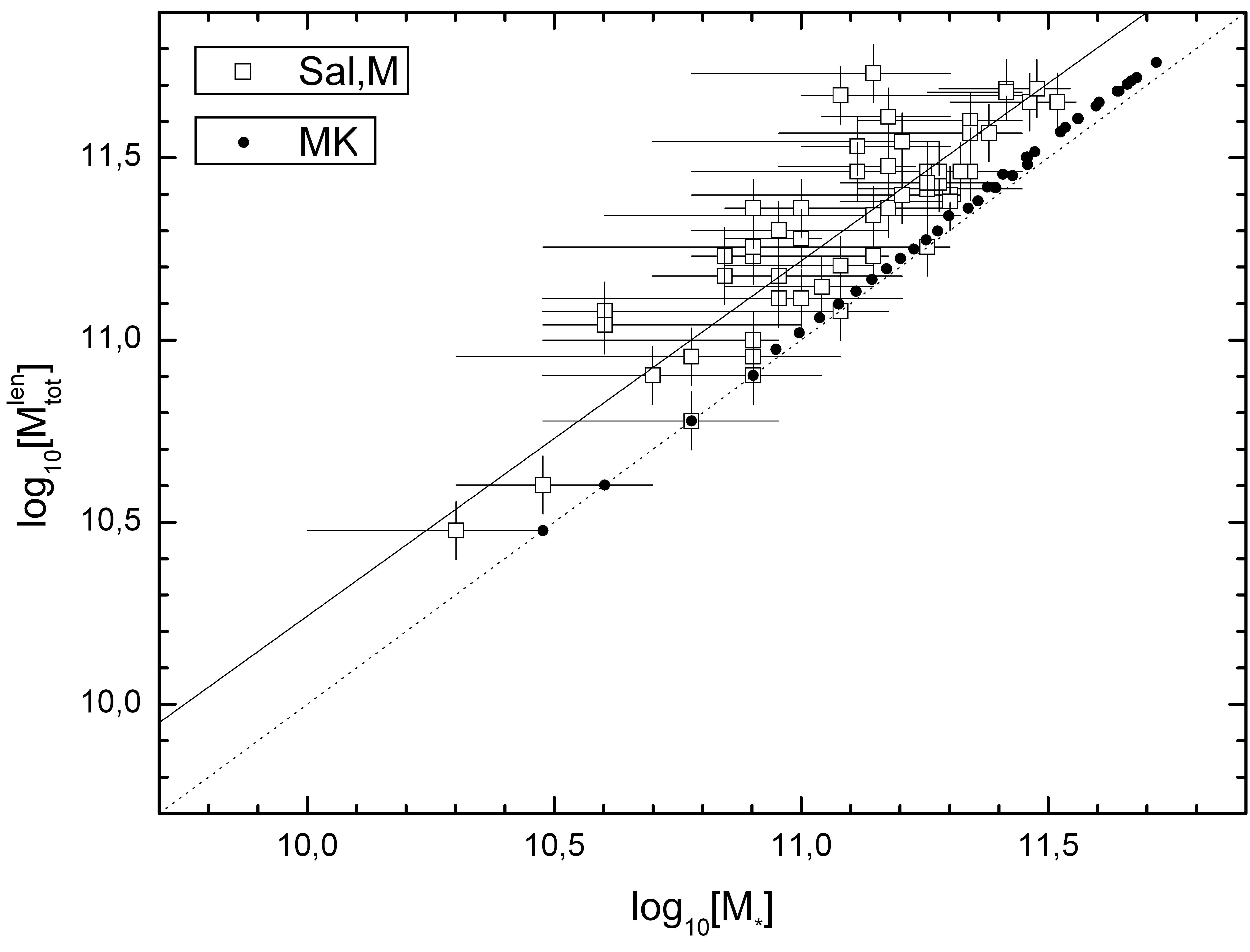}} b) \\
\end{minipage}
\caption{The linear plot (black dots) of observed total lens mass $M_{\text{tot%
}}^{\text{lens}}(\leq R_{\text{Ein}})$ for the $57$ lens galaxies of the
SLACS survey versus the luminous mass $M_{\ast }^{\text{MK}}(\leq R_{\text{%
Ein}})$ within the Einstein radius $R_{\text{Ein}}$obtained from our
algorithm [Table II]. The best-fit correlation line from different composite
stellar-population models [Bruzual and Charlot (BC) - panel (a) and Maraston (M) – panel (b)] and
IMFs [Salpeter (Sal)] and the one-to-one relation line $M_{\ast }/M_{\text{%
tot}}^{\text{lens}}=1$ are shown by solid and dotted lines respectively
(taken from [73]). Our linear plot and the best-fit correlation line are
parallel and can be merged into one another by a small constant numerical
shift.}
\end{figure}

\begin{figure}[ht]
\begin{minipage}[h]{0.55\linewidth}
\center{\includegraphics[width=1\linewidth]{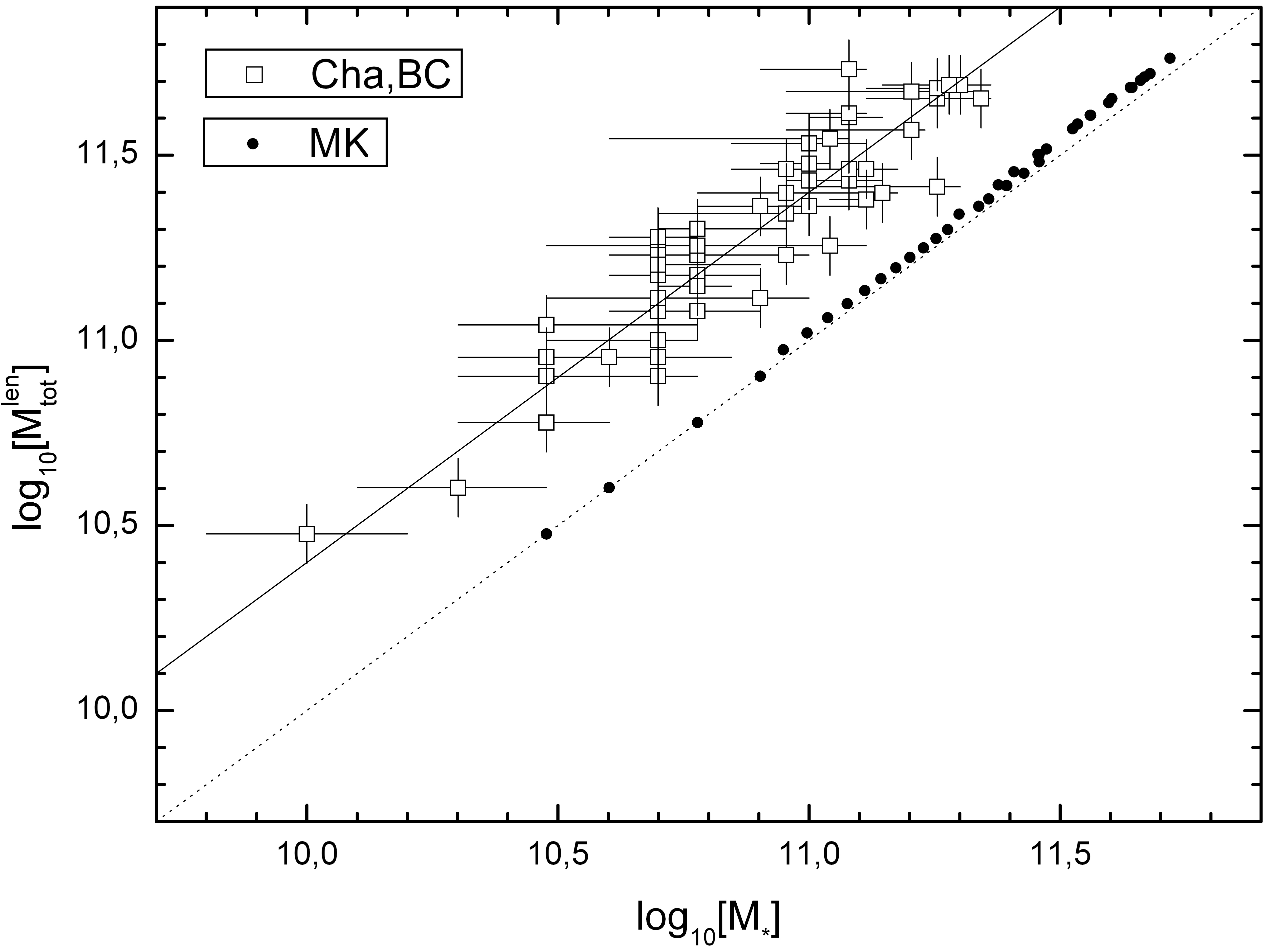}}  a) \\
\end{minipage}
\hfill
\begin{minipage}[h]{0.55\linewidth}
\center{\includegraphics[width=1\linewidth]{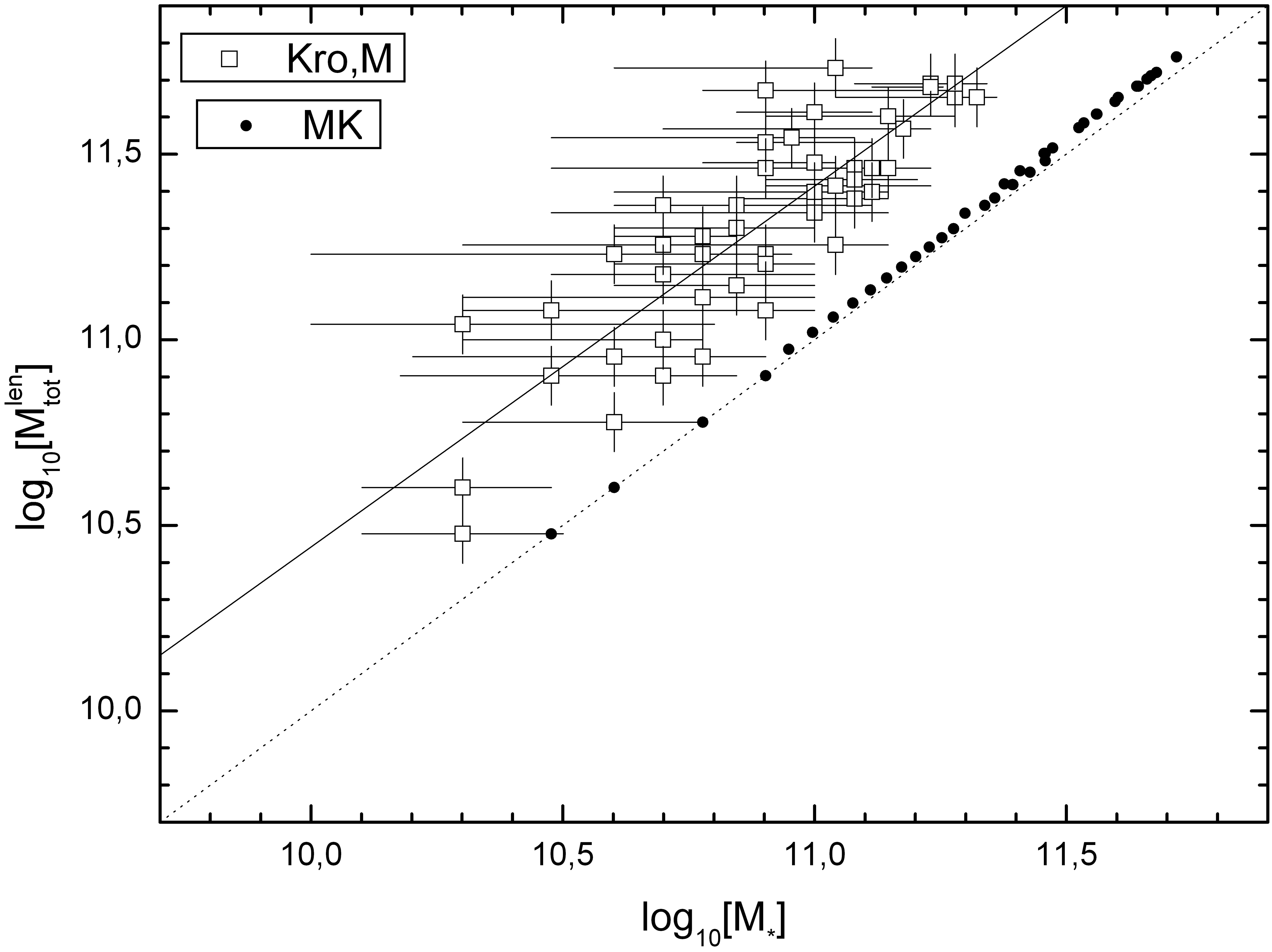}} b) \\
\end{minipage}
\caption{The linear plot (black dots) of observed total lens mass $M_{\text{tot%
}}^{\text{lens}}(\leq R_{\text{Ein}})$ for the $57$ lens galaxies of the
SLACS survey versus the luminous mass $M_{\ast }^{\text{MK}}(\leq R_{\text{%
Ein}})$ within the Einstein radius $R_{\text{Ein}}$obtained from our
algorithm [Table II]. The best-fit correlation line from different composite
stellar-population models [Bruzual and Charlot (BC)- panel (a)and Maraston (M)- panel (b)] and
IMFs [Chabrier (Cha), Kroupa (Kro)] and the one-to-one relation line $%
M_{\ast }/M_{\text{tot}}^{\text{lens}}=1$ are shown by solid and dotted
lines respectively (taken from [73]). Our linear plot and the best-fit
correlation line are parallel and can be merged into one another by a
constant but a bit larger numerical shift than in Fig.1, which indicates the
presence of comparatively more dark matter.}
\end{figure}

\section{Conclusions}

We started with an outline of Weyl conformal gravity focusing in particular
on the local and global spontaneous breakdown of conformal symmetry. Next,
we showed that the MK metric need not be an exclusive solution of conformal
gravity but can alternatively be viewed as a solution of a class of $f(R)$
gravity theories coupled to non-linear electrodynamic source. This
possibility endows our galactic mass decomposition scheme, and the derived
results, with more universality than thought heretofore.

We achieved two other goals in the foregoing work. First, we calculated
light deflection in the MK solution of Weyl conformal theory explicitly
bringing out the effect of the parameters $\gamma $ and $k$ appearing in the
metric (9). The calculation based on the vacuole method reveals that the
effect of $\gamma $ is to enhance two-way Schwarzschild bending ($4M_{\ast
}/b$) by an amount $+\gamma b$ (noting $R\simeq b$), while the effect of $k$
is to reduce it by an amount $-kbr_{b}$. The positive contribution $+\gamma
b $ is contrary to the previously obtained result $-\gamma b$ in the
literature [72,102,103]. Only for $\gamma <0$, the latter becomes positive
and truly imitates the effect of attractive dark matter but then the problem
is that the negative sign before $\gamma $ is \textit{opposite} to that
required to fit the rotation curves [72]. This long standing problem with
the MK solution has now been solved with the contribution $+\gamma b$ that
is positive for $\gamma >0$, the sign required to fit the rotation curves.
Notably, we required the same $\gamma >0$ for our mass decomposition as well
[See Eqs.(75) and (85)]. We argued in Sec.2 that the cause leading to
erroneous $-\gamma b$ lay in the illegitimate range of integration leading
to a metric signature change and an incomplete use of the Rindler-Ishak
method not respecting their prescribed positivity condition on $\left\vert
A\right\vert $.

The contribution $-kbr_{b}$ completely agrees with the expression obtained
by Ishak \textit{et al. }[69]. Let us for the moment notationally identify $%
k\equiv \Lambda /3$ so that $t_{\Lambda }=-\frac{\Lambda br_{b}}{3}$ and
recall a bit of curious history here. There has been a prevailing belief
that light deflection in the SdS spacetime is uninfluenced by the
cosmological constant $\Lambda $ appearing in the metric. The reason is that 
$\Lambda $ cancels out of the second order null geodesic equation as
probably first shown by Islam [71] $-$ naturally, the light bending
expression too does not contain it. This fact can be seen in the second
order null geodesic Eq.(35) from where the constant $k$ has disappeared,
even though $k$ does influence the non-null geodesic as seen in the exact
deflection Eq.(59). Eddington [122] was the first to examine one possible
manifestation of the latter in the perihelion advance of the planets and
derived a limit $\Lambda \leq 10^{-42}$ cm$^{-2}$. On the other hand, $%
\Lambda $ appears in the first order null geodesic equation, only its
further differentiation removes it from the second order equation.
Obviously, for the sake of logical consistency, the perturbative solution of
second order equation must also satisfy the first order equation, which
would then yield a relation among the involved constants, one of which is
the impact parameter $b$. An explicit calculation should in turn imply that
the removed $\Lambda $ would reappear in the light bending as well. This is
exactly what we have found happening $-$ a reflection of which is $%
t_{\Lambda }$ and the Sereno term $t_{\text{Sereno}}$ [See Eq.(70) and the
expression for $b$ derived in footnote 10].

Table I shows contributions to light bending coming from different factors
as well as the estimates of the average dark matter density $\left\langle
\rho \right\rangle _{\text{av}}^{\text{MK}}$ over the Einstein sphere. We
see from the last two columns that the positive contribution $t_{\gamma }$
is overtaken by the negative contribution $t_{k}$ in all the cases. However,
the combined effect is still one or two orders of magnitude less than the
contribution $t_{\text{Sch}}$. Only future measurement of higher order
corrections to $t_{\text{Sch}}$ could detect this combined effect, if any.
For the moment, the contributions in Eq.(71) to light deflection directly
impact the mass decomposition calculated in this paper.

Second, we applied the light deflection Eq.(71) together with a logical
input $\theta _{\text{Ein}}=\theta _{\text{Weyl}}$ to obtain the mass
decomposition into luminous and dark components of the lens galaxy within
its Einstein radius. The idea behind this input is that Weyl theory without
dark matter and Einstein theory with dark matter both should logically
predict exactly the \textit{same} numerical value for the angle of the
observed ring image of a background source if the former theory has to be in
the reckoning at all as a competing theory. This input automatically implies
that Weyl vacuum need not truly be a vacumm but an arena of cosmic
potentials $V_{\gamma _{0}}$ and $V_{k}$ bringing about quantitatively the
same lensing effect as would do the dark matter in Einstein's theory. In our
opinion, estimate of dark matter component thus provides an observable 
\textit{quantification} of the strength of such potentials symbolized by the
associated constants.

Table II shows mass decomposition in a representative subsample of $57$ lens
galaxies and that the observed total lens mass $M_{\text{tot}}^{\text{lens}}$
(luminous+dark) is linearly proportional to our derived luminous mass $%
M_{\ast }^{\text{MK}}$ across the subsample, which qualitatively agrees with
the existing conclusion in the literature as shown in Figs. 1 \& 2. Table II
provides the exact ratios from our analysis showing that those $57$ lens
galaxies are low in dark matter content within their Einstein radii.
Therefore, the ratios appear to be more akin to the simulation based on
(Sal,BC) than to others. Our ratios fall within the (Sal,BC) simulational
error bars for many individual galaxies and for a minority of cases the
ratios fall slightly outside the error bars in varying degrees. Since in the
literature [73] these ratios are also seen to vary rather significantly
depending on the choice of stellar population models and the IMFs, one could
equally regard the present analytic approach as yet another addition to the
existing scheme for decompositions.

Finally, it should be mentioned that there is some controversy about the
presence of dark matter halos around elliptical galaxies [123,124]. However,
massive ellipticals are generally considered as the result of fusion of
spiral galaxies. It is thus hard to understand how dark matter halos would
be present around spirals and absent after their fusion. The aim of the
present paper was however to focus on the inside of the Einstein radius and
not in the distant halo region.

\section{Acknowledgement}

Part of the work was supported by the Russian Foundation for Basic Research
(RFBR) under Grant No.16-32-00323.

\section{References}

[1] J. Oort, Bull. Astron. Inst. Netherl.\textbf{\ 6} (1931) 155.

[2] F. Zwicky, Helv. Phys. Acta \textbf{6} (1933) 110; Astrophys. J. \textbf{%
86} (1937) 217.

[3] J. Binney and S. Tremaine, \textit{Galactic dynamics}, Princeton
University Press, Princeton U.S.A. (1987).

[4] P. Salucci and M. Persic, Astron. Astrophys. \textbf{351} (1999) 442.

[5] Y. Sofue and V. Rubin, Ann. Rev. Astron. Astrophys. \textbf{39} (2001)
137.

[6] K.G. Begeman, A.H. Broeils and R.H. Sanders, Mon. Not. Roy. Astron. Soc. 
\textbf{249} (1991) 523.

[7] D.G. Barnes, R.L. Webster, R.W. Schmidt and A. Hughes, Mon. Not. Roy.
Astron. Soc. \textbf{309} (1999) 641.

[8] Y.-C. N. Cheng and L.M. Krauss, Astrophys. J. \textbf{514} (1999) 25.

[9] W.J.G. de Blok, S.S. McGaugh and V.C. Rubin, Astron. J. \textbf{122}
(2001) 2396.

[10] C.M. Trott and R.L. Webster, Mon. Not. Roy. Astron. Soc. \textbf{334}
(2002) 621.

[11] N.N. Weinberg and M. Kamionkowski, Mon. Not. Roy. Astron. Soc. \textbf{%
337} (2002) 1269.

[12] R.J. Smith, J.P. Blakeslee, J.R. Lucey and J. Tonry, Astrophys. J. 
\textbf{625} (2005) L103.

[13] T. Faber and M. Visser, Mon. Not. Roy. Astron. Soc. \textbf{372} (2006)
136.

[14] R.B. Metcalf and J. Silk, Phys. Rev. Lett. \textbf{98} (2007) 071302.

[15] S. Bharadwaj and S. Kar, Phys. Rev. D \textbf{68} (2003) 023516.

[16] M. Colpi, S.L. Shapiro and I. Wasserman, Phys. Rev. Lett. \textbf{57}
(1986) 2485.

[17] T. Matos, F.S. Guzm\'{a}n and D. Nu\~{n}ez, Phys. Rev. D \textbf{62}
(2000) 061301.

[18] P.J.E. Peebles, Phys. Rev. D \textbf{62} (2000) 023502.

[19] T. Matos and F.S. Guzm\'{a}n, Class. Quant. Grav. \textbf{17} (2000) L9.

[20] E.W. Mielke and F.E. Schunck, Phys. Rev. D \textbf{66} (2002) 023503.

[21] J.E. Lidsey, T. Matos and L.A. Ure\~{n}a-Lopez, Phys. Rev. D \textbf{66}
(2002) 023514.

[22] M.K. Mak and T. Harko, Phys. Rev. D\textbf{\ 70} (2004) 024010.

[23] K. Lake, Phys. Rev. Lett. \textbf{92} (2004) 051101.

[24] K.K. Nandi, I. Valitov and N.G. Migranov, Phys. Rev. D \textbf{80}
(2009) 047301; Erratum-\textit{ibid}. D \textbf{83} (2011)~029902.

[25] K. K. Nandi, A.I. Filippov, F. Rahaman, Saibal Ray, A. A. Usmani, M.
Kalam and A. DeBenedictis, Mon. Not. Roy. Astron. Soc.\textbf{\ 399} (2009)
2079.

[26] F. Rahaman, K.K. Nandi, A. Bhadra, M. Kalam and K. Chakraborty, Phys.
Lett. B \textbf{694} (2010) 10.

[27] A.A. Potapov, G.M.Garipova, and K.K. Nandi, Phys. Lett. B \textbf{753}
(2016)140.

[28] A.A. Usmani, F. Rahaman, S. Ray, K.K. Nandi, P.K.F. Kuhfittig, S.A.
Rakib and Z. Hasan, Phys.Lett. B \textbf{701} (2011) 388.

[29] S. Nojiri and S.D. Odintsov, Phys. Rept. \textbf{505} (2011) 59.

[30] O. Bertolami, C.G. Bohmer, T. Harko and F.S.N. Lobo, Phys. Rev. D 
\textbf{75} (2007) 104016.

[31] M. Bartelmann and R. Narayan, AIP Conf. Proc. \textbf{336} (1995) 307.

[32] A. Burkert, Astrophys. J. \textbf{447} (1995) L25.

[33] T. Harko and F.S.N. Lobo, Phys. Rev. D \textbf{83} (2011) 124051.

[34] U. Nucamendi, M. Salgado and D. Sudarsky, Phys. Rev. D \textbf{63}
(2001) 125016.

[35] J.F. Navarro, C.S. Frenk and S.D.M. White, Astrophys. J. \textbf{462}
(1996) 563; \textit{ibid.} \textbf{490} (1997) 493.

[36] S. Dodelson, E.I. Gates and M.S. Turner, Science \textbf{274} (1996) 69.

[37] J.-c. Hwang and H. Noh, Phys. Lett. B \textbf{680} (2009) 1.

[38] S.L. Dubovsky, P.G. Tinyakov and I.I. Tkachev, Phys. Rev. Lett. \textbf{%
94} (2005) 181102.

[39] J. Kluso\v{n}, S. Nojiri and S. D. Odintsov, Phys. Lett. B \textbf{726}
(2013) 918.

[40] S. Deser and G.W. Gibbons, Class. Quant. Grav. \textbf{15} (1998) L35.

[41] M. Ba\~{n}ados and P.G. Ferreira, Phys. Rev. Lett. \textbf{105} (2010)
011101.

[42] T. Delsate and J. Steinhoff, Phys. Rev. Lett. \textbf{109} (2012)
021101.

[43] P. Pani, V. Cardoso and T. Delsate, Phys. Rev. Lett. \textbf{107}
(2011) 031101.

[44] T. Delsate and J. Steinhoff, Phys. Rev. Lett. \textbf{109} (2012)
021101.

[45] X.-L. Du, K. Yang, X.-H. Meng and Y.-X. Liu, Phys. Rev. D \textbf{90}
(2014) 044054.

[46] T. Harko, F.S.N. Lobo, M.K. Mak and S.V. Sushkov, Phys. Rev. D\textbf{\
88} (2013) 044032

[47] A. Tamang, A.A. Potapov, R. Lukmanova, R. Izmailov and K.K. Nandi,
Class.Quant.Grav. \textbf{32} (2015) 235028.

[48] R. Izmailov, A.A. Potapov, A.I. Filippov, M. Ghosh and K.K. Nandi,
Mod. Phys. Lett. A\textbf{\ 30} (2015)1550056.

[49] A.A. Potapov, R. Izmailov, O. Mikolaychuk, N. Mikolaychuk, M. Ghosh,
K.K. Nandi, JCAP 07(2015) 018.

[50] M. Milgrom, Astrophys. J. \textbf{270} (1983) 365; \textit{ibid.} 
\textbf{270} (1983) 371; \textit{ibid.} \textbf{270} (1983) 384.

[51] M. Milgrom, Phys. Rev. Lett. \textbf{111}(2013) 041105.

[52] M. Milgrom, Phys. Rev. D \textbf{92} (2015) 044014.

[53] J. Bekenstein and M. Milgrom, Astrophys. J. \textbf{286} (1984) 7.

[54] R.A. Swaters, R.H. Sanders and S.S. McGaugh, Astrophys. J. \textbf{718}
(2010) 380.

[55] G. Gentile, B. Famaey, F. Combes, P. Kroupa, H.S. Zhao nad O. Tiret,
Astron. Astrophys. \textbf{472} (2007) L25.

[56] J.W. Moffat, JCAP 03 (2006) 004.

[56] G. Allemandi, A. Borowiec, M. Francaviglia and S.D. Odintsov, Phys.
Rev. D \textbf{72} (2005) 063505.

[58] S. Capozziello, V.F. Cardone and A. Troisi, Mon. Not. Roy. Astron. Soc. 
\textbf{375} (2007) 1423.

[59] \'{E}.\'{E}. Flanagan, Phys. Rev. D \textbf{74} (2006) 023002.

[60] P.D. Mannheim, Phys. Rev. D \textbf{75} (2007) 124006.

[61] P.D. Mannheim, Prog. Part. Nucl. Phys. \textbf{56} (2006) 340.

[62] P.D. Mannheim and D. Kazanas, Astrophys. J. \textbf{342} (1989) 635.

[63] V.A. Berezin, V.I. Dokuchaev and Yu.N. Eroshenko, Int. J. Mod. Phys. A 
\textbf{31} (2016) 1641004.

[64] S. Capozziello, V.F. Cardone and A. Troisi, JCAP 08 (2006) 001.

[65] P.D. Mannheim and J.G. O'Brien, Phys. Rev. Lett. \textbf{106} (2011)
121101.

[66] J.G. O'Brien and P.D. Mannheim, Mon. Not. R. Astron. Soc. \textbf{421}
(2012) 1273.

[67] K.K. Nandi and A. Bhadra, Phys. Rev. Lett. \textbf{109} (2012) 079001.

[68] A.S. Bolton, S. Burles, L.V.E. Koopmans \textit{et al.} Astrophys. J. 
\textbf{682} (2008) 964.

[69] M. Ishak, W. Rindler, J. Dossett, J. Moldenhauer and C. Allen, Mon.
Not. Roy. Astron. Soc. \textbf{388} (2008) 1279.

[70] W. Rindler and M. Ishak, Phys. Rev. D \textbf{76} (2007) 043006.

[71] J. N. Islam, Phys. Lett. A \textbf{97} (1983) 239.

[72] A. Edery and M. B. Paranjape, Phys. Rev. D \textbf{58 }(1998) 024011.

[73] C. Grillo, R. Gobat, M. Lombardi and P. Rosati, Astron. Astrophys. 
\textbf{501}, 461 (2009).

[74] G. 't Hooft, Int. J. Mod. Phys. D \textbf{24}, 1543001(2015)

[75] P.D. Mannheim, \textit{Living Without Supersymmetry -- the Conformal
Alternative and a Dynamical Higgs Boson}, arXiv:1506.01399 [hep-ph].

[76] A. Vilenkin, Phys. Lett. B \textbf{117 }(1982) 25.

[77] R. Penrose, Found. Phys. \textbf{44} (2014) 557.

[78] A. Zee, Annals of Phys. \textbf{151 }(1983) 431.

[79] Ya. B. Zel'dovich, JETP Lett. \textbf{9} (1970) 307.

[80] Ya. B. Zel'dovich and A. A. Starobinsky, JETP \textbf{34} (1972) 1159.

[81] P.D. Mannheim, Astrophys. J. \textbf{561} (2001) 1.

[82] P.D. Mannheim, Gen. Rel. Grav. \textbf{43} (2011) 703.

[83] R. Yang, B. Chen, H. Zhao, J. Li and Y. Liu, Phys. Lett. B \textbf{727}
(2013) 43.

[84] P.D. Mannheim, Phys. Rev. D \textbf{93} (2016) 068501.

[85] P.R. Phillips, Mon. Not. Roy. Astron. Soc. \textbf{448} (2015) 681.

[86] T.P. Sotiriou and V. Faraoni, Rev. Mod. Phys. \textbf{82} (2010) 451.

[87] M. Lubini, C. Tortora, J. Naf, Ph. Jetzer and S. Capozziello, Eur.
Phys. J. C \textbf{71 }(2011) 1834.

[88] S. Capozziello, M. De Laurentis and G. Lambiase, Phys.Lett. B \textbf{%
715} (2012) 1.

[89] S. Capozziello, S. Nojiri, S.D. Odintsov and A. Troisi, Phys. Lett. B 
\textbf{639} (2006)135.

[90] S. Capozziello, M. De Laurentis and O. Luongo, Int. J. Mod. Phys. D 
\textbf{24} (2014)1541002.

[91] S. Nojiri and S.D. Odintsov, Phys.Rev. D \textbf{78} (2008) 046006.

[92] S.D. Odintsov and V.K. Oikonomou, Phys. Rev. D 90 (2014) 124083.

[93] S. Nojiri and S.D. Odintsov, Phys. Rept. \textbf{505} (2011) 59.

[94] K. Bamba, S. Nojiri, S.D. Odintsov and D. S\'{a}ez-G\'{o}mez, Phys.
Rev. D \textbf{90 }(2014)124061.

[95] S. Nojiri and S.D. Odintsov, Gen. Rel. Grav. \textbf{36} (2004) 1765.

[96] S. Nojiri and S.D. Odintsov, Phys. Lett. B \textbf{735} (2014) 376.

[97] S. Capozziello and V. Faraoni, \textit{Beyond Einstein Gravity: A
Survey of Gravitational Theories for Cosmology and Astrophysics},
Fundamental Theories of Physics 170 (Springer, 2011).

[98] M.E. Rodrigues, J.C. Fabris, E.L.B. Junior and G.T. Marques, Eur.
Phys. J. C \textbf{76 }(2016) 250.

[99] S. Capozziello, A. Stabile and A. Troisi, Class.Quant.Grav. \textbf{25}
(2008) 085004.

[100] A.D. Dolgov and M. Kawasaki, Phys. Lett. B \textbf{573} (2003) 1.

[101] I. Bochicchio and E. Laserra, Int. J. Theor. Phys. \textbf{52} (2013)
3721.

[102] A. Bhattacharya, R. Isaev, M. Scalia, C. Cattani and K.K. Nandi, JCAP
09 (2010) 004.

[103] A. Bhattacharya, \ G.M. Garipova, E. Laserra, A. Bhadra and K.K.
Nandi, JCAP 02 (2011) 028.

[104] A. Einstein and E. Strauss, Rev. Mod. Phys. \textbf{17 }(1945)120;
Erratum: \textit{ibid.} \textbf{18 }(1946) 148.

[105] E. Schucking, Z. Phys. \textbf{137} (1954) 595.

[106] N.R. Sen, Ann. Phys. (Leipzig) \textbf{73} (1924) 365.

[107] C. Lanczos, Ann. Phys. (Leipzig) \textbf{74} (1924) 518.

[108] G. Darmois, \textit{Memorial de Sciences Mathematiques, Fasc XXV, 
Les Equations de la Gravitation. Einsteinienne}, (Gauthier-Villars, Paris, 1927), Chap. V

[109] W. Israel, Nuovo Cim. B \textbf{44 }(1966) 1; Erratum: \textit{ibid.} 
\textbf{48} (1967) 463.

[110] J. Bodenner and C.M. Will, Am. J. Phys. \textbf{71} (2003) 770.

[111] M. Sereno, Phys. Rev. Lett.\textbf{102} (2009) 021301.

[112] C. Cattani, M. Scalia, E. Laserra, I. Bochicchio and K. K. Nandi, Phys.
Rev. D \textbf{87} (2013) 047503.

[113] J.B. Hartle, \textit{Gravity:\ An Introduction to Einstein's General
Relativity}, Pearson Education, Inc., San Francisco (2003).

[114] M. Tegmark \textit{et al.} Phys. Rev. D\textbf{\ 69} (2004)103501);
B.W. Carroll and D.A. Ostlie, \textit{An Introduction to Modern Astrophysics}%
, 2nd Ed., Pearson Education, Inc., San Francisco (2007).

[115] Z. Horvath, L.A. Gergely, Z. Keresztes, T. Harko and F.S. N. Lobo,
Phys. Rev. D \textbf{84} (2011) 083006.

[116] G. Bruzual and S. Charlot, Mon. Not. R. Astron. Soc. \textbf{344}
(2003)1000.

[117] C. Maraston, Mon. Not. Roy. Astron. Soc. \textbf{362} (2005) 799.

[118] E.E. Salpeter, Astrophys. J. \textbf{121}, (1955) 161.

[119] G. Chabrier, Pub. Astron. Soc. Pacific \textbf{115} (2003) 763.

[120] P. Kroupa, Mon. Not. Roy. Astron. Soc. \textbf{322} (2001) 231.

[121] C. Grillo, Astrophys. J. \textbf{722} (2010) 779.

[122] A.S. Eddington, \textit{The Mathematical Theory of Relativity, }%
Cambridge University, Cambridge (1922), 8th Edition, 1960.

[123] A.J. Romanowsky \textit{et al}. Science \textbf{301} (2003) 1696.

[124] A. Dekel, F. Stoehr, G.A. Mamon, T.J. Cox, J.R. Primack, Nature
(London) \textbf{437} (2005) 707.

\end{document}